\documentclass[a4paper,12pt]{article}
\usepackage{jheppub} 
\usepackage{lineno}
\usepackage{hyperref}

\usepackage{amsmath,amsthm,amsfonts,amssymb,amscd}

\usepackage{caption}
\usepackage{fancyhdr}
\usepackage{mathrsfs}
\usepackage{physics}
\usepackage{tikz}
\usepackage{mathdots}
\usepackage{yhmath}
\usepackage{makecell}
\usepackage{cancel}
\usepackage{circuitikz}
\usepackage{color}
\usepackage{tikz-cd}
\usepackage{subcaption}
\usepackage{siunitx}
\usepackage{array}
\usepackage{placeins}
\usepackage{multirow}
\usepackage{gensymb}
\usepackage{tabularx}
\usepackage{extarrows}
\usepackage{booktabs}
\usetikzlibrary{fadings}
\usetikzlibrary{patterns}
\usetikzlibrary{shadows.blur}
\usetikzlibrary{shapes}
\usepackage{graphicx}

\title{\boldmath Flowery Horizons \& Bulk Observers: $sl^{(q)}(2,\mathbb{R})$ Drive in $2d$ Holographic CFT}

\author[a,b]{Jayashish Das,}
\emailAdd{jayashish.das[at]saha.ac.in}
\affiliation[a]{Saha Institute of Nuclear Physics, 1/AF Bidhannagar, Kolkata 700064, India}
\affiliation[b]{Homi Bhabha National Institute, Training School Complex, Anushaktinagar, Mumbai 400094, India.}

\author[a,b]{Arnab Kundu}
\emailAdd{arnab.kundu[at]saha.ac.in}

\abstract{We explore and analyze bulk geometric aspects corresponding to a driven two-dimensional holographic CFT, where the drive Hamiltonian is constructed from the $sl^{(q)}(2,\mathbb{R})$ generators. In particular, we demonstrate that starting with a thermal initial state, the evolution of the event horizon is characterized by distinct geometric transformations in the bulk which are associated to the conjugacy classes of the corresponding transformations on the CFT. Namely, the bulk evolution of the horizon is geometrically classified into an oscillatory (non-heating) behaviour, an exponentially growing  (heating) behaviour and a power-law growth with an angular rotation (the phase boundary), all as a function of the stroboscopic time. We also show that the explicit symmetry breaking of the drive is manifest in a flowery structure of the event horizon that displays a $U(1) \to {\mathbb Z}_q$ symmetry breaking. In the $q\to \infty$ limit, the $U(1)$ symmetry is effectively restored. Furthermore, by analyzing the integral curves generated by the asymptotic Killing vectors, we also demonstrate how the fixed points of these curves approximate a bulk Ryu-Takayanagi surface corresponding to a modular Hamiltonian for a sub-region in the CFT. Since the CFT modular Hamiltonian has an infinitely many in-equivalent extensions in the bulk, the fixed points of the integral curves can also lie outside the entanglement wedge of the CFT sub-region.}

\makeatletter
\gdef\@fpheader{}
\makeatother

\begin{document}
\maketitle
\begin{sloppypar}
\flushbottom

\section{Introductions}

Non-equilibrium dynamics is a challenging and important aspect to understand, especially since we do not know the organizing principles well enough that can provide us with a useful effective framework. See {\it e.g.}~\cite{Polkovnikov:2010yn} for a review on various aspects of physics out of equilibrium. It is therefore important to explore and understand soluble models that offer us enough control over the calculations. Conformal Field Theories (CFT), especially in two-dimensions, are indeed such a framework and we will extensively use the two-dimensional CFT in the present article.

Of particular importance to us are the driven systems, in which the Hamiltonian is endowed with a periodic protocol. In this framework several new physical phenomena emerge that have no equilibrium analogue, such as dynamical freezing\cite{PhysRevB.82.172402, PhysRevB.107.155117, Koch_2023, PhysRevB.90.174407, guo2024dynamicalfreezingexactlysolvable}, dynamical localization\cite{PhysRevB.89.165425, PhysRevB.95.014305, Baum_2018, Luitz_2017, PhysRevB.102.235114}, time-crystals\cite{Else_2020, Zaletel_2023, khemani2019briefhistorytimecrystals, PhysRevB.94.085112, PhysRevLett.118.030401, PhysRevLett.117.090402}, to name a few.

In this article, we will focus on periodic drives in two-dimensional CFTs with a large central charge.\footnote{This is equivalent to assuming that our CFT can be described by an AdS$_3$ dual geometry.} For a general CFT, using the conformal symmetry, one can already determine equal-time and un-equal time correlators of primary operators, entanglement entropy, stress-tensor expectation value as well as higher point functions such as the Out-of-time-order correlators (OTOCs).\footnote{Note that, while lower point correlation functions are completely fixed by the conformal invariance, higher point functions, such as the OTOCs, contain true dynamical features of the CFT. See {\it e.g.}~\cite{Biswas:2024mlq, Das:2019tga,Das:2021qsd,Das:2022jrr} for more details and explicit examples that emphasizes this point.} The drive Hamiltonian is constructed from the Virasoro generators in general, and in a particularly illuminating case, using the generators of the $sl(2,\mathbb{R})$ sub-algebra. In the $sl(2,\mathbb{R})$ case, the conjugacy classes of the corresponding algebra\footnote{Note that, the conjugacy classes of the $sl(2,\mathbb{R})$ algebra are essentially determined by the parameters of the drive Hamiltonian. There is an one-to-one but a non-linear relation between them.} determine the phase in the CFT: Namely, the hyperbolic, the parabolic and the elliptic conjugacy classes correspond to a heating, a phase-boundary and a non-heating phase in the CFT. By now, there is a large literature studying several aspects of such drives, see {\it e.g.}~\cite{Fan_2020, Fan_2021, Das_2021, PhysRevResearch.3.023044, das2022branedetectorsdynamicalphase, PhysRevResearch.2.023085, PhysRevB.97.184309, wen2018floquetconformalfieldtheory, Das_2024, das2024stretchedhorizonconformalfield, das2024notesheatingphasedynamics, PhysRevB.103.224303, PhysRevResearch.2.033461} and their holographic descriptions in\cite{das2022branedetectorsdynamicalphase, deBoer:2023lrd, MacCormack_2019, goto2021nonequilibratingblackholeinhomogeneous, Goto_2024, bernamonti2024boundaryinducedtransitionsmobiusquenches, kudlerflam2023bridgingquantumquenchproblems, Jiang:2024hgt}.

One can, nonetheless, consider a drive Hamiltonian constructed from $sl^{(q)}(2,\mathbb{R})$ generators. Therefore, the corresponding Hamiltonian now takes the form:
\begin{eqnarray}
    H = \alpha \left( L_0 + {\bar L}_0\right) + \beta\left(L_q + {\bar L}_q \right) + \gamma \left(L_{-q} + {\bar L}_{-q} \right) \ , \quad q > 1 \ . \label{Hfirst}
\end{eqnarray}
The generators $\{L_0, L_{\pm q}\}$ form a sub-algebra, which is usually denoted by $sl^{(q)}(2,\mathbb{R})$,  of the full Virasoro algebra and is therefore provide us with a tractable scenario for explicit computations of correlation functions and such. Unlike the $sl(2,\mathbb{R})$ algebra, there is no classification of its elements in terms of conjugacy classes. Instead now, one can define a quadratic Casimir of the $sl^{(q)}(2,\mathbb{R})$ algebra, which can be explicitly obtained as follows. One writes the Hamiltonian in (\ref{Hfirst}) as:
\begin{eqnarray}
   && H= \alpha L_0 + \beta^+ L_{q,+} + \beta^- L_{q,-} \ , \\
   &&  L_{q,+} = \frac{1}{2}\left( L_q + L_{-q}\right)  \ , \quad L_{q,+} = \frac{1}{2i}\left( L_q - L_{-q}\right) \ ,
\end{eqnarray}
which yields the quadratic Casimir to be: ${\cal C}_2 = - \alpha^2 + 4 \beta \gamma$. The sign of the Casimir now generates three possibilities that are in direct analogy with the conjugacy classes of $sl(2,\mathbb{R})$-algebra.

In this article, we focus on the geometric dual of this driven CFT, when the Hamiltonian is constructed from the generators of the $sl^{(q)}(2,\mathbb{R})$ algebra, with a particular focus on the behaviour of an initially existing event horizon in the bulk geometry, corresponding to the various phases. We will demonstrate how distinct stroboscopic evolution of the horizon emerges in the distinct phases.\footnote{Note that, it is easy to obtain a ``quench limit" from the driven framework, by setting the period of the drive to infinity. In this case, the corresponding horizon is a static one, nonetheless, the three possibilities are still realized.} Also the flower-like structure of horizons that we describe has independently appeared in a different physical scenario, namely in anisotropic quantum Hall droplets, highlighting a broader relevance of this geometric feature \cite{PhysRevX.14.011030}.

Before moving further, let us note that there are several complimentary ways of understanding or realizing a bulk geometric dual description of the driven CFT. First, and perhaps the most straightforward method makes direct use of the Bañados-geometries\cite{Ba_ados_1993, Ba_ados_1992}, which is the most general class of vacuum Einstein equations with a negative cosmological constant in three dimensions. In this case, given the data of the manifold on which the CFT is defined and the expectation value of the stress-tensor, one can immediately write down a metric in a particular coordinate basis that, under the Fefferman-Graham expansion reproduces the boundary CFT data. Even though this is starightforward, we will demonstrate that with the $sl^{(q)}(2,\mathbb{R})$ drives, the corresponding metric is rather involved to analyze, especially to understand the evolution of the event horizon. An additional problem is that the Bañados metric does not cover the entire bulk spacetime. There is a coordinate singularity where $\text{det}(g)=0$,  called the 'Wall' (see {\it e.g.}~\cite{Abajian:2023bqv,Abajian:2023jye}), with the horizon hidden within.

A related approach has been used recently in \cite{deBoer:2023lrd}, which we will heavily use in our present work. In this method, one simply maps the Bañados geometry to  the AdS-Poincaré patch and then to an AdS-Rindler. This is possible through the so-called Roberts' transformation which is known explicitly \cite{Roberts:2012aq}. The geometry beyond the wall is now accessed by this coordinate transformation. Once this is done, one simply tracks the evolution of the event horizon of the AdS-Rindler, through each coordinate transformation that is dual to the conformal transformation at the boundary. Starting with a thermal state, this method is best suited to visualize the stroboscopic evolution of the event horizon: In the heating phase, the evolution has an exponential growth, a power law growth on the phase transition line and an oscillatory behaviour in the non-heating phase. It is important to emphasize that all statements here are gauge-dependent in that these features are visible once a suitable coordinate patch is chosen. The full geometry is locally always an AdS$_3$, with boundary graviton excitations.

The gauge-dependent picture above naturally brings us to what a local bulk observer sees in the geometry. This was first analyzed in detail in \cite{das2022branedetectorsdynamicalphase}, and further explored in \cite{das2024notesheatingphasedynamics}. Here, one analyzes the integral curve equations that are generated by the bulk Killing vectors and the corresponding observer essentially flows along these integral curves.\footnote{Evidently, in the case of an $sl(2,\mathbb{R})$-drive, the bulk observer is described by global Killing vectors. However, for us, with $sl^{(q)}(2,\mathbb{R})$ drives, the corresponding Killing vectors obeys the Killing equations only in the asymptotic limit.}

In this article, we primarily analyze the horizon dynamics and the integral curves, following the frameworks of \cite{deBoer:2023lrd} and \cite{das2022branedetectorsdynamicalphase}, for $sl^{(q)}(2,\mathbb{R})$ valued drives and starting with an initial thermal state. In particular, we demonstrate that there is an explicit $U(1) \to {\mathbb Z}_q$ symmetry breaking under this drive, which is effectively restored in the $q\to \infty$ limit. Correspondingly, the event horizon in a suitable coordinate takes a flowery form with $q$ petals. Depending on the phase, the growth rate of the peaks of these petals are different: heating phase yields an exponentially growing peak of the petal, on the phase transition line, the peaks grow as a power law and in the non-heating phase the peaks oscillate. In all these cases, the growth occurs as a function of the stroboscopic time and along the radial direction of the bulk AdS$_3$ geometry. Except for the non-heating phase, in both heating phase as well as on the phase transition line, the peaks grow unbounded till they reach the conformal boundary of AdS. These points where they touch the conformal boundary correspond to the points on the CFT where stress-tensor expectation values also peak.

On the other hand, using the integral curves, we demonstrate the following: While we were unable to solve the integral curves in an analytic closed form, we analyzed the corresponding fixed points of these curves. Physically, these fixed points are associated with bulk observers with a vanishing acceleration.\footnote{Note that, in general, an integral curve describes a bulk observer with a non-trivial acceleration. Note that, as observed in \cite{das2024notesheatingphasedynamics}, the Floquent Hamiltonian can be mapped to the modular Hamiltonian of a sub-region in the CFT, where this map now contains explicit factors of $q$. Therefore, integral curves are not geodesics.} We observe that the fixed points, deep in the heating phase with $\alpha=0$, can be grouped into two categories: even $q$ and odd $q$. For even $q$, the fixed points of the integral curves correspond to points in the bulk geometry, whereas for odd $q$, they correspond to a co-dimension one hypersurfaces. As $q$ increases, for even $q$, the fixed points approach the conformal boundary and lie closer to the Ryu-Takayanagi surfaces corresponding to the modular Hamiltonian of the CFT. For odd $q$, the co-dimension one hypersurface itself approaches the Ryu-Takayanagi surface, near the conformal boundary. 

This article is divided into the following sections:
 In section $2$, we begin with some basic discussions on AdS$_3$ geometries, focussing in particular on the class of Bañados geometries, coordinate transformations and the basic framework in which we obtain the event horizon. We then discuss essential ingredients of the driven CFT framework in the next section. Section $4$ is dedicated towards fleshing out the choices of parameters in our problem that yield various phases and we use them to obtain and discuss the nature of the horizon in section $5$. Subsequently, in the next section we discuss the integral curves and their fixed points and finally we conclude. Several important technical details are summarized in appendices.

\label{sec:intro}
\section{\texorpdfstring{Basics Framework: $\text{AdS}_3 /\text{CFT}_2$}{}}

Let us begin with the basic ingredients that we will subsequently use heavily in this work. We will first review salient features of the geometric description.

\subsection{Bañados Geometries}

Let us consider a $\text{CFT}_2$ on a cylinder, ${\mathbb R}\times S^1$, with the Lorentzian metric: $ds^2=-dt^2+d\phi^2=-dx^+dx^-$. Here, $ \phi \backsim \phi+2\pi$ and $x^{\pm}=t\pm\phi$, where $t$ is the Lorentzian time of the CFT.\footnote{The corresponding Euclidean patch is evidently: $ds^2 = d\tau^2+d\phi^2$, which is obtained by $t \to - i \tau$.} For this CFT to have a holographic dual, we consider a large central charge and a sparse spectrum. The central charge is related to the Newton's constant:
$$
c=\frac{3}{2G_N} \gg 1 \,\,\,\,\,\, \Leftrightarrow \,\,\,\,\,\, G_N=\text{Newtons's Constant} \ll 1
$$
so that the geomtric dual is described by Einstein-gravity asymptotically  AdS\textsubscript{3} geometry (we set the AdS-radius to unity).

In three dimensions, the gravitational degrees of freedom are not dynamical, except for boundary gravitons. In an asymptotically AdS\textsubscript{3} geometry, these are described by the Brown–Henneaux boundary conditions\cite{Brown:1986nw,Terashima:2000gb},. The corresponding asymptotic symmetries were found to be two copies of the Virasoro algebra. Related to this, the most general solutions of this theory are given by the Bañados family of metrics:
\begin{align}
ds^2=\frac{dz^2-dx^+dx^-}{z^2}+\mathscr{L}^+(x^+)(dx^+)^2+\mathscr{L}^-(x^-)(dx^-)^2\notag \\-z^2\mathscr{L}^+(x^+)\mathscr{L}^-(x^-)dx^+dx^- \ . 
\label{eq:2.1}
\end{align}
Or, equivalently, by $z \xrightarrow{} \frac{1}{r} $,
\begin{equation}
ds^2=\frac{dr^2}{r^2}+\mathscr{L}^+(x^+)(dx^+)^2+\mathscr{L}^-(x^-)(dx^-)^2-\left(r^2+\frac{\mathscr{L}^+(x^+)\mathscr{L}^-(x^-)}{r^2} \right)dx^+dx^- \ .
\label{eq:2.2}
\end{equation}
Note that the metric \eqref{eq:2.1} is an exact solution to vacuum  AdS\textsubscript{3} Einstein equations.\footnote{The general solution in (\ref{eq:2.2}) can be viewed as the Fefferman-Graham expansion, which, in the three-dimensional case, truncates.} The (anti-) holomorphic functions $\mathscr L^{\pm}(x^{\pm})$ are assumed to be smooth and exhibit 2$\pi$-periodicity in their arguments.

Given the general class of geometries in (\ref{eq:2.2}), different choices for $\mathscr L^{\pm}(x^{\pm})$ corresponds to different geometries which are asymptotically $\text{AdS}_3$ \cite{Ecker:2019ocp,Mandal:2014wfa} .
\begin{center}

    Poincaré Patch \quad $\Longrightarrow \quad \mathscr L^{\pm}(x^{\pm}) =0$ \ ,
    \\
    Global  \quad $\text{AdS}_3 \Longrightarrow \quad \mathscr L^{\pm}(x^{\pm}) =-\frac{1}{4}$ \ ,
    \\
    Family of non-extremal BTZ black holes \quad $\Longrightarrow \mathcal \quad \mathscr L^{\pm}(x^{\pm}) =$Possitive Constants \ .

\end{center}
In general the holomorphic and the anti-holomorphic functions are given by
\begin{equation}
    \mathscr L^{\pm}(x^{\pm})=-\frac{1}{2}\text{Sch}\{F_{\pm}(x^{\pm}),x^{\pm}\} \ , 
    \label{eq:2.3}
\end{equation}
where $\text{Sch}\{F(x),x\}$ is Schwarzian derivative given by 
\begin{equation}
    \text{Sch}\{F(x),x\}=\frac{F'''}{F'}-\frac{3}{2}\left(\frac{F''}{F'} \right)^2 \ .
    \label{eq:2.4}
\end{equation}
Locally, the family of geometries in \eqref{eq:2.2} can be transformed into the Poincaré patch AdS$_3$:
\begin{equation}
    ds^2=\frac{-dT^2+dX^2+dZ^2}{Z^2}\,\,\,\,\, , \,\,\,\,\, Z>0 \ ,
\label{eq:2.5}
\end{equation}
where $X^{\pm}=X\pm T$, can be obtained by the Robert's Transformations:
\begin{align}
    X^+ & =F_-(x^-)-\frac{2F_-'(x^-)^2F_+''(x^+)}{4r^2F_-'(x^-)F_+'(x^+)+F_-''(x^-)F_+''(x^+)} \notag \ ,
    \\
    X^- &=F_+(x^+)-\frac{2F_+'(x^+)^2F_-''(x^-)}{4r^2F_-'(x^-)F_+'(x^+)+F_-''(x^-)F_+''(x^+)} \notag \ , 
    \\
    Z & =\frac{4r(F_-'(x^-)F_+'(x^+))^\frac{3}{2}}{4r^2F_-'(x^-)F_+'(x^+)+F_-''(x^-)F_+''(x^+)} \ .
    \label{eq:2.6}
\end{align}
These maps provide an explicit realization of uniformization in CFT. To be more precise, given a coordinate transformations $x^+ \xrightarrow{} F_+(x^+)$ and $x^-\xrightarrow{} F_-(x^-) $, we are able to ``geometrize" a quantum state  $\ket{\phi } $ , such that the expectation value of stress tensor is \cite{Caputa:2022zsr}:
\begin{align}
    \bra{\phi}T_{\pm\pm}(x^+)\ket{\phi} & =-\frac{c}{24\pi}\text{Sch}\{F_{\pm}(x^{\pm}),x^{\pm}\} \ . 
    \label{eq:2.7}
\end{align}
We immediately obtain the dual gravity geometry that determines the functions from \eqref{eq:2.3}:
\begin{equation}
    2\pi \bra{\phi}T_{\pm\pm}\ket{\phi}=\frac{c}{6}\mathscr L^{\pm}(x^{\pm}) \ .
    \label{eq:2.8}
\end{equation}
Therefore, given the CFT data $\{\gamma_{ab}, \langle T_{ab} \rangle\}$, where $\gamma_{ab}$ is the manifold on which the CFT is defined and $\{\langle T_{ab} \rangle\}$ is the expectation value of the CFT stress-tensor, we are able to obtain an explicit metric. While the metric is locally AdS$_3$, it may be represented in a particular coordinate system that is suitable for a particular class of observers. Therefore, the corresponding physical phenomena associated to this observer will be described by the corresponding coordinate patch. It may, nonetheless, be useful to map this patch into the Poincaré patch using the Roberts' transformations. We will make heavy use of this in this work.

\subsection{Event Horizon of Bañados geometries}

Let us now write the Bañados metric \eqref{eq:2.2} in terms of $t$ and $\phi$ coordinates:
\begin{multline}
     ds^2=\frac{dr^2}{r^2}+\Big[-r^2-\frac{\mathscr{L}^+\mathscr{L}^-}{r^2}\Big]dt^2+\Big[r^2+\frac{\mathscr{L}^+\mathscr{L}^-}{r^2}\Big]d\phi^2
     \\
     +\mathscr{L}^+(dt^2+d\phi^2+2d\phi dt)+\mathscr{L}^-(dt^2+d\phi^2-2d\phi dt) \ .
     \label{eq:2.9}
   \end{multline}
Suppose, now, $\mathscr L^{\pm}(x^{\pm})$ is constant: $\mathscr L^{+}(x^{+})=\mathscr L^{-}(x^{-})=\mathscr L$\,\, then the metric becomes:
\begin{equation*}
    ds^2=\frac{dr^2}{r^2}+\Big[r^2+\frac{\mathscr{L}}{r^2}+2\mathscr{L}\Big]d\phi^2-\Big[r^2+\frac{\mathscr{L}}{r^2}-2\mathscr{L}\Big]dt^2 \ ,
\end{equation*}
\begin{equation}
  \implies  ds^2=\frac{dr^2}{r^2}+\Big[r+\frac{\mathscr{L}}{r}\Big]^2d\phi^2-\Big[r-\frac{\mathscr{L}}{r}\Big]^2dt^2 \ . 
  \label{eq:2.10}
\end{equation}
Now we make a new coordinate transformation \cite{Sheikh-Jabbari:2014nya}:
$$
\rho=r+\frac{\mathscr{L}}{r} \ .
$$%
The metric becomes:
\begin{equation}
    ds^2=\frac{d\rho^2}{\rho^2-4\mathscr{L}}+\rho^2d\phi^2-(\rho^2-4\mathscr{L})dt^2 \ . 
    \label{eq:2.11}
\end{equation}
For positive constants value of $\mathscr L$ this metric is the standard BTZ Black Hole metric.

To find the event horizon we can set $g_{tt}$ to zero and we get:
\begin{equation}
    \rho_{H}=2\sqrt{\mathscr{L}} \ , 
    \label{eq:2.12}
\end{equation}
so in the ($r,t,\phi$) coordinate it is at $r_H=\sqrt{\mathscr L}$. The BTZ metric is given by
\begin{equation}
    ds^2=\frac{d\rho^2}{\rho^2-\rho_+^2}+\rho^2d\phi^2-(\rho^2-\rho_+^2)dt^2 \ . 
    \label{eq:2.13}
\end{equation}
where $\rho_+$ is the radius of the BTZ Black Hole, with an inverse temperature $ \beta_{\rm{th}} $ given by $\rho_+=\frac{2\pi}{\beta_{\rm{th}}}$ (correspondingly, $r_+=\frac{\pi}{\beta_{\rm{th}}}$).

Now, let us consider the general case, when $\mathcal L^{\pm}(x^{\pm})$ depends on $x^{\pm}$. We can now map the Bañados metric to the Poincaré $\text{AdS}_3$ metric and determine the location of the event horizon. The Poincaré metric is given by
\begin{equation}
    ds^2=\frac{-dT^2+dX^2+dZ^2}{Z^2} \ .
    \label{eq:2.14}
\end{equation}
The Poincaré coordinates $(T,X,Z)$ can be used to define an auxiliary coordinate system, which is given below:\footnote{Note that, these are essentially the embedding space coordinates.} 
\begin{align}
x^0 &=\frac{T}{Z} \ , \quad  x^1 =\frac{Z^2+X^2-T^2-1}{2Z} \notag
\\
x^2 &=\frac{X}{Z} \ , \quad x^3 =\frac{Z^2+X^2-T^2+1}{2Z} \ .
\label{eq:2.15}
 \end{align}
Now we can explicitly write down the the $\text{AdS}_3$ metric in Rindler coordinates \cite{Parikh:2012kg}:
\begin{equation}
    ds^2=-\xi^2 dt_R^2+\frac{d\xi^2}{1+\xi^2}+(1+\xi^2)d\chi^2 \ ,
    \label{eq:2.16}
\end{equation}
where the relation to the Poincaré patch is given by
\begin{align}
x^0 =\xi \sinh{t_R} \ , \quad  x^1 &=\xi \cosh{t_R} \ ,\notag \\
x^2 =\sqrt{1+\xi^2}\sinh{\chi} \ , \quad x^3 &=\sqrt{1+\xi^2}\cosh{\chi} \ .
\label{eq:2.17}
 \end{align}
In these coordinates the event horizon is at $\xi=0$. Using \eqref{eq:2.15} and \eqref{eq:2.17} imposing $\xi=0$ we get:
\begin{equation}
X^2+Z^2=1 \,\,\,\,\,\, , \,\,\,\,\,\, T=0 \ .
\label{eq:2.18}
\end{equation}
So, the Rindler horizon is given by a Semi-Circle in the Poincaré $\text{AdS}_3$.

Now, given a general Bañados metric, one can map it to the Poincaré patch using the  Roberts's Transformation. Subsequently, the semi-circle in the Poincaré patch can be studied, which corresponds to the Rindler horizon in AdS. As we will explicitly demonstrate, this horizon will display salient features of the driven system. These steps are summarized in appendix \ref{appenC}. Here we give a diagramatic representation. 
\vspace{8pt}

\tikzset{every picture/.style={line width=0.75pt}} 

\begin{tikzpicture}[x=0.75pt,y=0.75pt,yscale=-1,xscale=1]
 
\draw    (137,61) -- (196.2,61.12) -- (237.2,61.2) ;
\draw [shift={(239.2,61.2)}, rotate = 180.11] [color={rgb, 255:red, 0; green, 0; blue, 0 }  ][line width=0.75]    (10.93,-3.29) .. controls (6.95,-1.4) and (3.31,-0.3) .. (0,0) .. controls (3.31,0.3) and (6.95,1.4) .. (10.93,3.29)   ;
\draw    (335,62) -- (427.2,62.2) ;
\draw [shift={(429.2,62.2)}, rotate = 180.12] [color={rgb, 255:red, 0; green, 0; blue, 0 }  ][line width=0.75]    (10.93,-3.29) .. controls (6.95,-1.4) and (3.31,-0.3) .. (0,0) .. controls (3.31,0.3) and (6.95,1.4) .. (10.93,3.29)   ; 
\draw    (471,84) -- (471.19,142.2) ;
\draw [shift={(471.2,144.2)}, rotate = 269.81] [color={rgb, 255:red, 0; green, 0; blue, 0 }  ][line width=0.75]    (10.93,-3.29) .. controls (6.95,-1.4) and (3.31,-0.3) .. (0,0) .. controls (3.31,0.3) and (6.95,1.4) .. (10.93,3.29)   ;
\draw    (378,159) -- (266.2,159.2) ;
\draw [shift={(264.2,159.2)}, rotate = 359.9] [color={rgb, 255:red, 0; green, 0; blue, 0 }  ][line width=0.75]    (10.93,-3.29) .. controls (6.95,-1.4) and (3.31,-0.3) .. (0,0) .. controls (3.31,0.3) and (6.95,1.4) .. (10.93,3.29)   ;

\draw (24,54) node [anchor=north west][inner sep=0.75pt]   [align=left] {{\fontfamily{helvet}\selectfont Bañados Metric}};
\draw (160,40) node [anchor=north west][inner sep=0.75pt]   [align=left] {{\fontfamily{helvet}\selectfont {\small Robert's}}};
\draw (139,64) node [anchor=north west][inner sep=0.75pt]   [align=left] {{\fontfamily{helvet}\selectfont {\small Transformation}}};
\draw (249,44) node [anchor=north west][inner sep=0.75pt]   [align=left] {{\fontfamily{helvet}\selectfont  \ Poincar\'{e} }\\{\fontfamily{helvet}\selectfont AdS Metric}};
\draw (346,40) node [anchor=north west][inner sep=0.75pt]   [align=left] {{\fontfamily{helvet}\selectfont {\small Embedding}}};
\draw (339,63) node [anchor=north west][inner sep=0.75pt]   [align=left] {{\small {\fontfamily{helvet}\selectfont Coordinates}}};
\draw (436,43) node [anchor=north west][inner sep=0.75pt]   [align=left] {{\fontfamily{helvet}\selectfont  \ \ Rindler }\\{\fontfamily{helvet}\selectfont AdS Metric}};
\draw (408,93) node [anchor=north west][inner sep=0.75pt]   [align=left] {{\small {\fontfamily{helvet}\selectfont Horizon}}\\{\small {\fontfamily{helvet}\selectfont at $\xi=0$}}};
\draw (381,149) node [anchor=north west][inner sep=0.75pt]   [align=left] {{\fontfamily{helvet}\selectfont  \ Horizon in Poincar\'{e}}};
\draw (282,139) node [anchor=north west][inner sep=0.75pt]   [align=left] {{\small {\fontfamily{helvet}\selectfont Semi-Circle's}}};
\draw (294,162) node [anchor=north west][inner sep=0.75pt]   [align=left] {{\fontfamily{helvet}\selectfont {\small Equation}}};
\draw (144,150) node [anchor=north west][inner sep=0.75pt]   [align=left] {{\fontfamily{helvet}\selectfont Bañados Horizon}};

\end{tikzpicture}

\section{Driven 2D CFTs}

Given a particular drive protocol, the evolution operator is simply obtained as:
\begin{eqnarray}
U\left( nT, 0\right) &  = & {\mathcal T} \left\{ {\rm exp} \left[- i n \int_0^T H(t') dt' \right] \right\} \nonumber\\
& = & {\rm exp} \left[ - i n T H_F(T)\right] \ , 
\label{eq:3.1}
\end{eqnarray}
where $H(t)$ is the actual time-dependent Hamiltonian and $H_F$ is the Floquet Hamiltonian, $n$ is the number of periodic drives with a time-period $T$. When $H(t)$ is constructed from the generators of a given algebra, $H_F$ can clearly be obtained as a general linear combination of the generators of the algebra. Here, we will study the dynamics as a function of the stroboscopic time $n$.

\subsection{The set-up}

We consider a $(1+1)d$ CFT on a cylinder of topology ${\mathbb R}\times S^1$, with  periodic boundary condition. The cylinder has a total circumference of $L$. At this point, the details of the CFT data are not relevant. We can choose the initial state to be the vacuum state $\ket{0}$ of the usual CFT Hamiltonian (which we denote by $H_0$) or a thermal state $\rho=e^{-\beta_{\rm th} H_0}$ with inverse temperature $\beta_{\rm th}$.

We will choose a periodic and time-dependent Hamiltonian with which the CFT will evolve. This time-dependence can be represented by choosing a particular protocol for the drive. For example: 
\\[0.5cm]
\textbf{Protocol-I}
\begin{center}
$H(t) = \begin{cases}
 H_0\,\, , & \mbox{if } 0<t\le T_0 \hspace{60pt}T_0 +T_1=T, \text{Total Time period} \\
 H_1 \,\,,  & \mbox{if } T_0\le t \le T
\end{cases}
$
\end{center}
\tikzset{every picture/.style={line width=0.75pt}} 
\begin{tikzpicture}[x=0.75pt,y=0.75pt,yscale=-1,xscale=1]
\draw    (100,127) -- (156,127) ;
\draw    (156,127) -- (156,74.61) ;
\draw    (156,74.61) -- (208,74.61) ;
\draw    (208,125.8) -- (208,74.61) ;
\draw    (208,125.8) -- (265,125.8) ;
\draw    (265,125.8) -- (264,74.61) ;
\draw    (264,74.61) -- (320,74.61) ;
\draw    (320,125.8) -- (320,74.61) ;
\draw    (320,125.8) -- (376,125.8) ;
\draw    (376,125.8) -- (375,74.61) ;
\draw    (375,74.61) -- (428,74.61) ;
\draw    (428,125.8) -- (428,74.61) ;
\draw    (428,125.8) -- (484,125.8) ;
\draw    (484,125.8) -- (483,74.61) ;
\draw    (483,74.61) -- (539,74.61) ;
\draw   (98,136.9) -- (112,139) -- (112,137.95) -- (140,137.95) -- (140,139) -- (154,136.9) -- (140,134.8) -- (140,135.85) -- (112,135.85) -- (112,134.8) -- cycle ;
\draw   (154,136.9) -- (168,139) -- (168,137.95) -- (196,137.95) -- (196,139) -- (210,136.9) -- (196,134.8) -- (196,135.85) -- (168,135.85) -- (168,134.8) -- cycle ;
\draw    (290,156.97) -- (366,156.97) ;
\draw [shift={(368,157)}, rotate = 180.88] [color={rgb, 255:red, 0; green, 0; blue, 0 }  ][line width=0.75]    (10.93,-3.29) .. controls (6.95,-1.4) and (3.31,-0.3) .. (0,0) .. controls (3.31,0.3) and (6.95,1.4) .. (10.93,3.29)   ;
\draw    (539,74.41) -- (540,125.8) ;
\draw    (540,125.8) -- (554,125.8) ;
\draw    (568,125.8) -- (582,125.8) ;
\draw    (594,125.8) -- (610,125.8) ;

\draw (109,104) node [anchor=north west][inner sep=0.75pt]   [align=left] {$\displaystyle H_0$};
\draw (171,49) node [anchor=north west][inner sep=0.75pt]   [align=left] {$H_1$};
\draw (113,147) node [anchor=north west][inner sep=0.75pt]   [align=left] {$T_0$};
\draw (169,147) node [anchor=north west][inner sep=0.75pt]   [align=left] {$T_1$};
\draw (325,157) node [anchor=north west][inner sep=0.75pt]   [align=left] {t};

\end{tikzpicture}
\\[1cm]
\textbf{Protocol-II}
\begin{center}
$H(t) = \begin{cases}
 H_1\,\, , & \mbox{if } 0<t\le T_1 \hspace{60pt}T_1 +T_0=T, \text{Total Time period} \\
 H_0 \,\,,  & \mbox{if } T_1\le t \le T
\end{cases}
$
\end{center}
\tikzset{every picture/.style={line width=0.75pt}} 
\begin{tikzpicture}[x=0.75pt,y=0.75pt,yscale=-1,xscale=1]
%

\draw    (156,74.61) -- (208,74.61) ;
\draw    (208,125.8) -- (208,74.61) ;
\draw    (208,125.8) -- (265,125.8) ;
\draw    (265,125.8) -- (264,74.61) ;
\draw    (264,74.61) -- (320,74.61) ;
\draw    (320,125.8) -- (320,74.61) ;
\draw    (320,125.8) -- (376,125.8) ;
\draw    (376,125.8) -- (375,74.61) ;
\draw    (375,74.61) -- (428,74.61) ;
\draw    (428,125.8) -- (428,74.61) ;
\draw    (428,125.8) -- (484,125.8) ;
\draw    (484,125.8) -- (483,74.61) ;
\draw    (483,74.61) -- (539,74.61) ;
\draw   (210,136.9) -- (224,139) -- (224,137.95) -- (252,137.95) -- (252,139) -- (266,136.9) -- (252,134.8) -- (252,135.85) -- (224,135.85) -- (224,134.8) -- cycle ;
\draw   (154,136.9) -- (168,139) -- (168,137.95) -- (196,137.95) -- (196,139) -- (210,136.9) -- (196,134.8) -- (196,135.85) -- (168,135.85) -- (168,134.8) -- cycle ;
\draw    (390,156.97) -- (466,156.97) ;
\draw [shift={(468,157)}, rotate = 180.88] [color={rgb, 255:red, 0; green, 0; blue, 0 }  ][line width=0.75]    (10.93,-3.29) .. controls (6.95,-1.4) and (3.31,-0.3) .. (0,0) .. controls (3.31,0.3) and (6.95,1.4) .. (10.93,3.29)   ;
\draw    (539,74.41) -- (540,125.8) ;
\draw    (540,125.8) -- (554,125.8) ;
\draw    (568,125.8) -- (582,125.8) ;
\draw    (594,125.8) -- (610,125.8) ;

\draw (221,100) node [anchor=north west][inner sep=0.75pt]   [align=left] {$\displaystyle H_0$};
\draw (171,49) node [anchor=north west][inner sep=0.75pt]   [align=left] {$H_1$};
\draw (225,147) node [anchor=north west][inner sep=0.75pt]   [align=left] {$T_0$};
\draw (169,147) node [anchor=north west][inner sep=0.75pt]   [align=left] {$T_1$};
\draw (425,157) node [anchor=north west][inner sep=0.75pt]   [align=left] {t};

\end{tikzpicture}
\\[1cm]
The above protocol can be summarily captured by the following Hamiltonian:
\begin{equation}
    H_\theta=\int_0 ^Ldx\Big[1-\tanh(2\theta)\cos (q\frac{2\pi x}{L})\Big]T_{00}(x)
 \,\,,\,\,   q=2,3,4....   ,   \theta>0
 \label{eq:3.2}
\end{equation}
Here, the original Hamiltonian is $H_0=H_{\theta=0}$ and the deformed Hamiltonian is $H_1=H_{\theta \neq0}$.

The system\footnote{This simply means that in the Schr\"{o}dinger picture the initial CFT state evolves, whereas in the Heisenberg picture CFT operators evolve. Because of the state-operator correspondence, we can think about the evolution of either the state or the operator.} evolves periodically with a period $T$ under the periodic drive protocol, therefore the entire evolution occurs in steps ($n$) of $T$, {\it i.e.}~for a total time of $t = n T$. We will study the dynamical features as a function of the discrete time-step, $n$, otherwise known as the stroboscopic time.

The full Hamiltonian can be written in terms of $\{L_0,L_{\pm q}\}$ generators, which form the corresponding algebra: $sl^{(q)}(2,\mathbb{R})$.
\begin{equation}
H_{\rm deform}=\frac{2\pi}{L}\Big[L_0-\tanh(2\theta)\frac{L_q+L_{-q}}{2}-\frac{c}{24} \Big]+\text{anti-chiral parts} \ . 
\label{eq:3.3}
\end{equation}
In general, $q$ is any positive integer. In the case $q=1$, we have the standard $sl(2,\mathbb{R})$ algebra, which upon exponentiation also forms the $SL(2,\mathbb{R})$ group. In this case, the $sl(2,\mathbb{R})$ algebra can be classified in terms of its conjugacy classes: hyperbolic, parabolic and elliptic.\footnote{It is useful to note that in terms of conformal transformations, hyperbolic transformation corresponds to dilatation transformation, parabolic elements correspond to the shear transformation and elliptic elements correspond to rotation transformations.}

In this work, we will mainly consider $q > 1$, in which case the $sl^{(q)}(2,\mathbb{R})$ algebra does not have a conjugacy classification. Instead, in this case, one can obtain a natural classification in terms of the signature of the quadratic Casimir. Unlike the $q=1$ case, the vacuum state will evolve under for $q>1$ and therefore the vacuum stress-tensor expectation value already yields a non-zero and non-trivial response under the drive\cite{wen2018floquetconformalfieldtheory}.

We will focus on the Lorentzian time evolution of driven CFT.
However, we will first work in the Euclidean patch, where the time-evolution is captured by standard sequences of conformal transformations on the plane, and subsequently performing an analytic continuation to the Lorentzian patch. This will simply be a sequence of $sl^{(q)}(2,\mathbb{R})$ transformations on the plane, fused together.

\subsection{Operator Evolution \& Phases}

The Euclidean cylinder coordinates are:
\begin{equation}
    w=\tau +i x \,\, , \,\, \Bar{w}=\tau-i x \ .
    \label{eq:3.4}
\end{equation}
We will use $\tau$ and $t$  to denote imaginary and real time respectively and the spatial variable $x$ is periodically identified with a period of $2\pi $ . The analytic continuation is given by $\tau\xrightarrow{}i t$. We will calculate the operator evolution in the Heisenberg picture: ${\cal O}(\tau) = e^{\tau H} {\cal O}(0) e^{- \tau H}$. So, we begin with a primary operator on an initial $t=0$ time-slice (therefore, $\tau=0$)\footnote{As we are only giving the deformation in the spatial direction.} with $w=i x$ and $\Bar{w}=-i x$ and subsequently time evolve it using the appropriate Hamiltonian.

Now to explicitly calculate the operator evolution after a single-cycle driving, we consider the conformal mapping from the cylinder to plane geometry:
$$
z=e^{\left(\frac{2\pi q w}{L} \right)}=e^{(\frac{2\pi w}{l})} \ , \quad w=\tau+i x \ , \quad l = \frac{L}{q} \ .
$$
For a fixed $\tau$ , $z$ winds $q$-times as $x$ increases from $0$ to $L$. This means that the $z$ coordinate describes a $q$-sheeted Riemann surface.

On the Riemann surface the operator evolution becomes a dilatation\cite{wen2018floquetconformalfieldtheory} i.e. now the time-evolution is realized by simply applying conformal transformations on the $q$-sheeted Riemann sheet. One can find the operator evolves as
\begin{equation}
    e^{H^{(z)}\tau}\mathcal{O}(z,\bar{z})e^{-H^{(z)}\tau}=\left(\frac{\partial z_1}{\partial z}\right)^h \left(\frac{\partial \bar{z_1}}{\partial \bar{z}}\right)^{\bar{h}}\mathcal{O}(z_1,\bar{z}_1) \ . \label{eq:3.5}
\end{equation}
For a single-cycle drive, the operator evolution on the $q$-sheeted Riemann surface is described by this M\"{o}bius transformation:
$$
z_1=\frac{\alpha z+\beta}{\gamma z+\delta} \,\,\,\,\,\,\, , \,\,\,\,\,\,\,
 M(\tau)=\begin{bmatrix}
\alpha & \beta \\
\gamma & \delta
\end{bmatrix}\,\,\,
\in \,\,\,SL(2,\mathbb{C}) \ . 
$$
After imposing the condition $\alpha\delta-\beta\gamma=1$ and performing an analytic continuation $\tau=i t$ we get:\footnote{Note that, as groups, $SU(1,1) \cong SL(2,\mathbb{R})$, and both are subgroups of $SL(2,\mathbb{C})$.}
$$ 
z_1=\frac{az+b}{b^{\ast}z+a^{*}} \,\,\,\,\,\,\, , \,\,\,\,\,\,\, M(\tau \to it)=\begin{bmatrix}
a & b \\
b^{*} & a^{*}
\end{bmatrix}\,\,\,
\in \,\,\,SU(1,1)\,\,\, \text{with} \,\,|a|^2-|b|^2=1 \ .
$$
Here the various parameters are given by \cite{Wen:2022pyj}
$$
a=\cos(\frac{\pi T_\theta}{l_{\rm eff}})+i\cosh(2\theta)\sin(\frac{\pi T_\theta}{l_{\rm eff}})\,\,\,\, , \,\,\,\, b=-i\sinh(2\theta)\sin(\frac{\pi T_\theta}{l_{\rm eff}}) \ ,
$$
$$
 \text{with}\,\,\,\, l_{\rm eff}=l\cosh(2\theta)  \,\, , \,\, l=\frac{L}{q} \,\, , \,\, T_{\theta=0}=T_0 \,\, , \,\, T_{\theta\neq 0}=T_1 \,\, , \,\, 
$$
Here, $l_{\rm eff}=l\cosh(2\theta)$ emerges as an effective length of the total system.

We will denote the M\"{o}bius transformation matrices corresponding to $H_0$ and $H_1$ Hamiltonian by $M_0$ and $M_1$, respectively and compose them to together to obtain results for a single cycle. Under the transformation applied $n$ times, $z$ maps to $z_n$ in the complex plane and hence $w$ to $w_n$. On the other hand, $z_n$ is connected to $z$ by a M\"{o}bius transformation. So After $n$-drive cycles we have:
\begin{equation}
    z_n=\frac{a_n z+b_n}{b_n^* z+a_n^*} \,\,\,\,\,,\,\,\,\,\,\, z_n=e^{\frac{2\pi w_n}{l}} \ ,
    \label{eq:3.6}
\end{equation}
so that:
\begin{equation}
    w_n=\frac{l}{2\pi}\ln{\frac{a_n e^{\frac{2\pi w}{l}}+b_n}{b_n^* e^{\frac{2\pi w}{l}}+a_n^*}} \ . 
    \label{eq:3.7}
\end{equation}
Then the stress tensor which is not a primary operator transforms as:
\begin{align}
  &  T_{\rm cyl}(w)=\left(\frac{dz}{dw}\right)^2\left[ T_{\rm Plane}(z)-\frac{c}{12}\text{Sch}\{w,z\} \right] \ , 
    \label{eq:3.8}
    \\
& \implies   \langle T_{\rm cyl}(w) \rangle=\frac{\pi^2 c}{6\beta_{\rm th}^2} \notag \ .
\end{align}
Now, the vacuum expectation value of the chiral part of the Energy-Momentum tensor density is given by
\begin{align}
  &  \langle T(w,n) \rangle= \left(\frac{\partial w_n}{\partial w} \right)^2 \langle T(w_n) \rangle+\frac{c}{12}\text{Sch}\{w_n,w\} \ , \notag
    \\
    &=\left(\frac{\partial w_n}{\partial w} \right)^2 \frac{\pi^2 c}{6\beta_{\rm th}^2} +\frac{c}{12}\text{Sch}\{w_n,w\} \ .
    \label{eq:3.9}
\end{align}
In terms of the number of the drive cycles, $n$, this yields:
\begin{equation}
    \frac{1}{2\pi} \langle T(w,n) \rangle =\frac{-q^2\pi c}{12L^2}+\frac{\pi c}{12L^2}.(q^2-1).\frac{1}{|a_n e^{\frac{2\pi  w}{l}}+b_n|^4 } \ ,
     \label{eq:3.10}
\end{equation}
and
\begin{equation}
   \frac{1}{2\pi}  \langle \Bar{T}(\Bar{w},n) \rangle =\frac{-q^2\pi c}{12L^2}+\frac{\pi c}{12L^2}.(q^2-1).\frac{1}{|a_n e^{\frac{2\pi  \Bar{w}}{l}}+b_n|^4 } \ .
     \label{eq:3.11}
\end{equation}
Here, the first term is the Casimir energy and $a_n \, , b_n$ are the elements of the $SU(1,1)$ matrix after $n$-drive cycles. The total energy can be obtained by integrating \eqref{eq:3.10} w.r.t $x$ from $0$ to $L$ and it is given by
\begin{equation}
    E(n)=\frac{\pi c}{6L}(q^2-1)(|a_n|^2+|b_n|^2)-\frac{q \pi c}{6L} \ .
    \label{eq:3.12}
\end{equation}
Also, note that:
\begin{equation}
    |\text{Tr}(M_0 M_1)^n|=2 \left|\cosh{2\theta}\sin{\frac{\pi T_1}{l_{\rm eff}}}\sin{\frac{\pi T_0}{l}}-\cos{\frac{\pi T_1}{l_{\rm eff}}}\cos{\frac{\pi T_0}{l}} \right| \ . 
    \label{eq:3.13}
\end{equation}
Such driven CFTs exhibit three different behaviours, commonly used classifier is the Trace of the Evolution matrix \cite{Han:2020kwp}, given $M=(M_0 M_1)$:
\begin{align*}
|\text{Tr}(M)| & >2 \xrightarrow \,\,\,\text{Heating Phase} \\
|\text{Tr}(M)| & =2 \xrightarrow \,\,\, \text{Phase transition} \\
|\text{Tr}(M)| & <2 \xrightarrow \,\,\, \text{Non-Heating Phase}
\end{align*}
We will discuss the Holographic descriptions, focusing on the dynamics of the corresponding event horizons, associated to these different phases.

If, on the other hand, we begin with a thermal state, the expectation value of the stress tensor from \eqref{eq:3.9} will be:
\begin{equation}
    \langle T(w,n) \rangle=\frac{-q^2\pi^2 c}{6L^2}+\left(\frac{q^2\pi^2 c}{6L^2}+\frac{\pi^2 c}{6\beta_{\rm th}^2}\right).\frac{1}{\left|a_n e^{\frac{2\pi w}{l}}+b_n \right|^4 } \ . 
    \label{eq:3.14}
\end{equation}
So, the initial thermal state is described by $a_n=i$ and $b_n=0$, which gives back the thermal energy density $E(x)=\frac{\pi^2 c}{6\beta_{\rm th}^2}$.

Now the total energy becomes:
\begin{equation}
    E(n)= \left(\frac{\pi c}{6l^2}+\frac{\pi c}{6\beta^2} \right)l(|a_n|^2+|b_n|^2)-\frac{\pi c}{6l} \ .
    \label{eq:3.15}
\end{equation}
In Euclidean picture, $w$ changes to $w_n$ under the drive. In the Holographic description, the evolution of the CFT states can be expressed in terms of the action of a bulk diffeomorphism $f$, especially large gauge transformations, on the initial state. This leads to a new state, by exciting boundary gravitons dynamically, as a function of the number of drive cycles \cite{Choo:2022lgm}. Schematically:
\begin{equation}
    x_n=f_{t,n}(x) \ . 
    \label{eq:3.16}
\end{equation}
The corresponding CFT stress-tensor expectation value becomes:
\begin{equation}
    \langle T_{\pm \pm} \rangle =f_t^{\prime}(x_{\pm})^2 \langle T \rangle_{\beta_{\rm th}}-\frac{c}{24\pi}\text{Sch}\{f_t({x_{\pm}}),x_{\pm}\} \ . 
    \label{eq:3.17}
\end{equation}
Where $\langle T \rangle_{\beta_{\rm th}}=\frac{\pi c}{12 \beta_{\rm th}^2}$. Note that here the sign before the Schwarzian changes compared to \eqref{eq:3.9} the reason is due to the coordinate transformation $w \to ix$ and the Schwarzian term contains two un-cancelled derivatives \cite{Fan_2021}.

Now, calculating $\frac{\partial w_n}{\partial w}$ and then substituting $w$ by $i x$, we obtain the function $f_{t,n}^{\prime}(x)$. Upon integration we can find the desired diffeomorphism $f_{t,n}(x)$. We will now use these diffeomorhphism to map the Bañados geometry to the Poincar\'{e} patch and subsequently to the AdS-Rindler coordinate. This will provide us with an explicit set of coordinates in which the AdS-Rindler event horizon can be translated back to the Bañados coordinates. We will then study the dependence of this event horizon, as a function of the number of the drive cycle or the so-called stroboscopic time.

\section{Different Phases of Driven CFT}

Before discussing the dynamics of the associated event-horizon, let us briefly review the parametric regimes which correspond to the various phases in the CFT. Evidently, in the Holographic description, we implicitly assume that the CFT has a large central charge and a dual description.

\subsection{Heating Phase}

Let us choose: $\Big(\frac{T_0}{l},\frac{T_1} {l_{\rm eff}}\Big)=\Big(\frac{1}{2},\frac{1}{2}\Big)$. At this point the CFT exhibits the heating phase for arbitrary $\theta \neq 0$ as $|\text{Tr}(M_0 M_1)|=2\cosh{2\theta} $. The evolution matrix for the usual CFT Hamiltonian is:
$$
M_0=\begin{bmatrix}
i & 0 \\
0 & -i
\end{bmatrix} \ , 
$$
while, for the deformed case, it is:
$$
M_1=\begin{bmatrix}
i\cosh{2\theta} & -i\sinh{2\theta} \\
i\sinh{2\theta} & -i\cosh{2\theta}
\end{bmatrix} \ .
$$
Now, using protocol-I we have:
$$
(M_0 M_1)=\begin{bmatrix}
-\cosh{2\theta} & \sinh{2\theta} \\
\sinh{2\theta} & -\cosh{2\theta}
\end{bmatrix} \ .
$$
The $n$-th evolution is obtained by multiplying products of $(M_0 M_1)$, $n$ times, which yields:
$$
U(nT\, , \, 0)=(M_0M_1)^n=(-1)^n\begin{bmatrix}
\cosh(2n\theta) & -\sinh(2n\theta)\\
-\sinh(2n\theta) & \cosh(2n\theta)
\end{bmatrix} \ . 
$$
Therefore: $a_n=(-1)^n\cosh(2n\theta)$ and $b_n=-(-1)^n\sinh(2n\theta)$.

To obtain the gravity dual description, we replace the $w$ and $\Bar{w}$ coordinates by $x^{\pm}$. The latter are used to write down the Bañados geometries. 

Therefore:
$
|a_n e^{\frac{2\pi w}{l}}+b_n|^4 = \left [\cosh(4n \theta)-\cos{\frac{2\pi  x^{\pm}}{l}} \sinh(4n\theta) \right]^2 \ . 
$
The stress tensor in heating phase is given by,
\begin{equation}
   \langle T_{\pm\pm}(x^{\pm}) \rangle+\frac{q^2\pi c}{12L^2}=\frac{\pi c}{12L^2}(q^2-1)\left[\cosh(4n \theta)-\cos{\frac{2\pi x^{\pm}}{l}} \sinh(4n\theta) \right]^{-2} \ , 
    \end{equation}
    \label{eq:4.1}
    and 
    \begin{align}
         {\langle T_{\pm\pm}(x^{\pm}) \rangle}_{\beta_{\rm th}} +\frac{q^2\pi c}{12L^2}=\left(\frac{\pi c}{12L^2}q^2+\frac{\pi c}{12\beta_{\rm th}^2}\right) \left[\cosh(4n \theta) -\cos{\frac{2\pi x^{\pm}}{l}} \sinh(4n\theta) \right]^{-2} \ . 
         \label{eq:4.2}
    \end{align}
    Hence the energy is: 
    \begin{equation}
    E(n)= \left(\frac{\pi c}{6l^2}+\frac{\pi c}{6\beta^2}\right)l\left(\cosh^2({2n\theta})+\sinh^2({2n\theta}) \right)-\frac{\pi c}{6l} \ .
    \notag
    \end{equation}
\begin{equation}
\implies   E(n)= \left(\frac{\pi c}{6l^2}+\frac{\pi c}{6\beta^2}\right)l\cosh({4n\theta}) -\frac{\pi c}{6l} \ .
\label{eq:4.3}
\end{equation}
In this heating phase the one-point function shows exponential decay (for large enough $n$, $\cosh(4n\theta) \sim e^{4n\theta} \sim \sinh(4n\theta)$) and the total energy shows an exponential growth.

\label{phasetran_line}
\subsection{Phase Transition Line}

In this case, $\text{Tr}|(M_0 M_1)|=2$ , so from \eqref{eq:3.13} we can choose: $\frac{\pi T_0}{l}=\frac{\pi}{2}$ and $\cosh{2\theta}\sin{\frac{\pi T_1}{l_{\rm eff}}}=1$. So the Phase transition occurs at:
\begin{eqnarray}
\left(\frac{T_0}{l},\frac{T_1} {l_{\rm eff}}\right) =\left(\frac{1}{2},\frac{1}{\pi}\sin^{-1}\left({\frac{1}{\cosh{2\theta}}}\right)\right) \ . 
\label{eq:4.4}
\end{eqnarray}
As before, the transformation corresponding to the standard CFT Hamiltonian is:
$$
M_0=\begin{bmatrix}
i & 0 \\
0 & -i
\end{bmatrix} \ , 
$$
and for the deformed Hamiltonian, it is:
$
M_1=\begin{bmatrix}
\tanh{2\theta}+i & -i\tanh{2\theta} \\
\tanh{2\theta} & \tanh{2\theta}-i
\end{bmatrix} \ . 
$
The $n$-th evolution is obtained by multiplying products of $(M_0 M_1)$ , $n$-times and we obtain:
$$
U(nT\, , \, 0)=(M_0M_1)^n=(-1)^n\begin{bmatrix}
1-in\tanh{2\theta} & -n\tanh{2\theta} \\
-n\tanh{2\theta} & in\tanh{2\theta}+1
\end{bmatrix} \ . 
$$
As before, we can directly replace the $w$ and $\Bar{w}$ coordinates by the Bañados coordinates $x^{\pm}$ . 

Therefore:
$
|a_n e^{\frac{2\pi w}{l}}+b_n|^4 =\Big[1-2n\tanh{2\theta}\Big[\cos{\frac{2\pi  x^{\pm}}{l}}+n(\sin{\frac{2\pi  x^{\pm}}{l}}-1)\tanh{2\theta}\Big]\Big]^2 \ . 
$
Hence the stress tensor expectation value is:
\begin{align}
   {\langle T_{\pm\pm}(x^{\pm}) \rangle}_{\beta_{\rm th}}+\frac{q^2\pi c}{12L^2}=\left(\frac{\pi c}{12L^2}q^2+\frac{\pi c}{12\beta^2} \right)\Big[1-2n\tanh{2\theta}\Big[\cos{\frac{2\pi  x^{\pm}}{l}}+\notag \\ n \left(\sin{\frac{2\pi  x^{\pm}}{l}}-1 \right)\tanh{2\theta}\Big]\Big]^{-2} \ .
   \label{eq:4.5}
    \end{align} 
    Finally, the energy is:
    \begin{equation}
    E(n)= \left(\frac{\pi c}{6l^2}+\frac{\pi c}{6\beta^2} \right)l\left(1+2n^2\tanh({2n \theta})\right)-\frac{\pi c}{6l} \ .
    \label{eq:4.6}
     \end{equation}
On the phase transition line, the one-point function shows the power-law decay and the total energy exhibits a power law growth.

\subsection{Non-Heating Phase}

Let us now choose: $ \Big(\frac{T_0}{l},\frac{T_1} {l_{\rm eff}}\Big)=\Big(1,\frac{1}{2}\Big)$ The evolution matrices are:
\begin{center}

$M_0=\begin{bmatrix}
-1 & 0 \\
0 & -1
\end{bmatrix} $ 
and $M_1=\begin{bmatrix}
i\cosh{2\theta} & -i\sinh{2\theta} \\
i\sinh{2\theta} & -i\cosh{2\theta}
\end{bmatrix} $ \ .
\end{center}
\begin{center}

$(M_0M_1)=\begin{bmatrix}
-i \cosh{2\theta} & i \sinh{2\theta} \\
-i \sinh{2\theta} & i \cosh{2\theta}
\end{bmatrix} $
and $(M_0M_1)^2=\begin{bmatrix}
-1 & 0 \\
0 & -1
\end{bmatrix} $\ .  
\end{center}
We can clearly see the trace of evolution matrix is less than 2.

The full evolution matrix $(M_0 M_1)^n $ depends on whether $n$ is even or odd; for even $n$, it becomes a constant matrix and for odd $n$ it becomes a function of $\phi$ .  In the Non-heating phase both the stress-tensor one-point function and the total energy oscillate. The total energy is given by
\begin{eqnarray}
    && E(n)= \left(\frac{\pi c}{6l^2}+\frac{\pi c}{6\beta^2} \right)l\left(\cosh^2({2n\theta})+\sinh^2({2n\theta}) \right)-\frac{\pi c}{6l} \quad \text{for odd n} \ ,\label{eq:4.7} \\
    && \hspace*{40mm} E(n)=\frac{\pi^2 c}{6\beta_{\rm th}^2}  \quad \text{for even n} \ . \label{eq:4.8}
\end{eqnarray}

{\bf A Useful Note:}
Before moving further, let us take a quick stock of the three distinct cases in terms of the time-periods for which the deformed Hamiltonian acts on the system.
\begin{center}

    Heating Phase \quad $\Longrightarrow \quad \Big(\frac{T_0}{l},\frac{T_1} {l_{\rm eff}}\Big)=\Big(\frac{1}{2},\frac{1}{2}\Big)$ \ ,
    \\
    Phase Transition  \quad $\Longrightarrow \quad \left(\frac{T_0}{l},\frac{T_1} {l_{\rm eff}}\right) =\left(\frac{1}{2},\frac{1}{\pi}\sin^{-1}\left({\frac{1}{\cosh{2\theta}}}\right)\right) $ \ ,
    \\
    Non-heating Phase \quad $\Longrightarrow \Big(\frac{T_0}{l},\frac{T_1} {l_{\rm eff}}\Big)=\Big(1,\frac{1}{2}\Big) $ \ .

\end{center}
From this we can clearly see that there is an order on the time-scale during which the deformation is turned on, associated to each of the three possibilities:
\begin{center}
    $(T_1)_{\text{heating}} \geq (T_1)_{\text{phase transition}} \geq (T_1)_{\text{non-heating}} \ ,$
\end{center}
where $T_0$ is kept fixed. Therefore, keeping the deformation beyond a critical time-scale results in the heating phase. Note that, this critical point does not depend on the initial state of the CFT.\footnote{Note that there is a parametric regime in the $T_0/l$ vs $T_1/l$ space that corresponds to each of the three possibilities. We have explicitly chosen representative points corresponding to each of these \cite{Fan_2020,PhysRevResearch.3.023044}.}

\section{Dual Geometries of Driven CFT}

We will now discuss how the event horizon dynamically evolves in the Holographic dual description. We will only keep track of the coordinate transformations and the associated evolution of the event horizon, as this will already be sufficiently illuminating. For the three different cases, let us discuss the evolution of the horizon separately. For convenience, we will begin with a thermal state, instead of the vacuum state.

\subsection{Heating Phase}

We will now make explicit use of the diffeomorphisms discussed in section $2$. For example, first, using \eqref{eq:2.8} and \eqref{eq:4.2} we obtain the following function:
\begin{equation}
    \mathscr{L}^{\pm}(x^{\pm})=-\frac{q^2\pi^2 }{L^2}+\left(\frac{q^2\pi^2 }{L^2}+\frac{\pi^2 }{\beta^2}\right)\frac{1}{[\cosh(4n \theta)-\cos{\frac{2\pi x^{\pm}}{l}} \sinh(4n\theta)]^2} \ .
    \label{eq:5.1}
\end{equation}
From this we can clearly see the functional form of $\mathscr{L^{\pm}}$ are the same. Furthermore, we are calculating the expectation value of the stress tensor on $t$-constant slices, so that we have $\mathscr{L}(x^+)=\mathscr{L}(x^-)$.

Now we can find the diffeomorphisms $f_{\pm}(x^{\pm})$ using \eqref{eq:2.3}. We define:
\begin{align}
   & F_{\pm}(x^{\pm})=\tanh\left({\sqrt{\frac{12\pi \langle T \rangle_{\beta_{\rm{th}}}}{c}}f_{n,\pm}(x^{\pm})}\right) \ ,  \label{eq:5.2}
    \\    
    & \implies F_{\pm}(x^{\pm}) =\tanh\left({\frac{\pi}{\beta_{\rm th}}f_{n,\pm}(x^{\pm})}\right) \ .
    \label{eq:5.3}
\end{align}
such that the stress tensor expectation value becomes:
\begin{equation}
    {\langle T_{\pm\pm}(x^{\pm}) \rangle}_{\beta_{\rm th}}=-\frac{c}{24\pi}\text{Sch}\{F_{\pm}(x^{\pm}),x^{\pm}\} \ . 
    \label{eq:5.4}
\end{equation}

Now, in general, we have:
\begin{equation}
\frac{\partial w_n}{\partial w}\xrightarrow{w=ix} f_n^{\prime}(x)=\frac{1}{\cosh (4 n \theta)-\cos (x) \sinh (4 n \theta)} \ .
\label{eq:5.5}
\end{equation}
Integrating the above equation w.r.t $x$ we obtain:
\begin{equation}
    f_n(x)=\frac{L \tan ^{-1}\left(e^{4 n \theta  } \tan \left(\frac{\pi  q x}{L}\right)\right)}{\pi  q} \ . 
    \label{eq:5.6}
\end{equation}
Note that, this diffeomorphism is naturally characterized by the stroboscopic time parameter $n$, and can therefore be directly used to study the stroboscopic time-dependence of the corresponding geometry.

We can now determine the event horizon by mapping the Bañados geometry to the Poincar\'{e} patch and then identifying the semi-circle equation in Poincaré coordinates. Thus, using \eqref{eq:2.6} and \eqref{eq:2.18}, we obtain the location of the  horizon to be:
\begin{equation}
    r(x^-)=\frac{\pi f'_{n}(x^-)}{\beta_{\rm th}}\sqrt{1-\frac{\beta_{\rm th}^2 f''_{n}(x^-)^2}{4 \pi^2 f'_{n}(x^-)^4}} \ . 
    \label{eq:5.7}
\end{equation}
We have used: $\mathscr{L}(x^+)=\mathscr{L}(x^-)$, which implies $ f_{n,+}(x^{+})= f_{n,-}(x^{-})=f_{n}$. Now,  using equation (\ref{eq:5.5}), the horizon is given by
\begin{equation}
      r(x^-)=\frac{\sqrt{4 \pi ^2-\beta_{\rm th} ^2 q^2 \sinh ^2(4 n \theta  ) \sin ^2(q x^-)}}{2 \beta_{\rm th}  \cosh (4 n \theta  )-2 \beta_{\rm th}  \sinh (4 n \theta  ) \cos (q x^-)} \ . 
      \label{eq:5.8}
\end{equation}
We have set $L=2\pi$ above, for simplicity.

We will now study equation (\ref{eq:5.7}) in detail. Note that because of the presence of the square-root in the denominator, setting $\beta_{\rm th} \to \infty$ does not yield any real solution for $r(x^-)$: {\it i.e.}~there is no event horizon if we begin with a vanishing temperature vacuum state. This clearly justifies our choice of keeping a finite $\beta_{\rm th}$\footnote{We also show this explicitly in Appendix-\ref{appenC}.}. On the other hand, in the $\beta_{\rm th} \to 0$ limit, there is always a real event horizon.

Secondly, keeping $q$ and $\theta$ fixed, it is clear that an event horizon exists as long as the $\sin(q x^-) \sinh(4 n\theta)$ does not exceed an upper bound. This fact endows an angular dependence to the event horizon.\footnote{There is no $t$ dependence, since we are considering $t=0$ slices.} Interestingly, for large enough $n$, because of a conspiracy between the numerator and the denominator, the location of the horizon actually grows exponentially with $n$. This is the hallmark of the heating phase. If we set $x^-=0,$ that is we only focus on how the peak of the horizon grows, we get $r(x^-)=\frac{\pi}{\beta_{\rm th}}e^{4n\theta}$ , which exhibits an exponential growth.

\begin{figure}[h!tbp]  
    \centering
    
    \includegraphics[width=\linewidth]{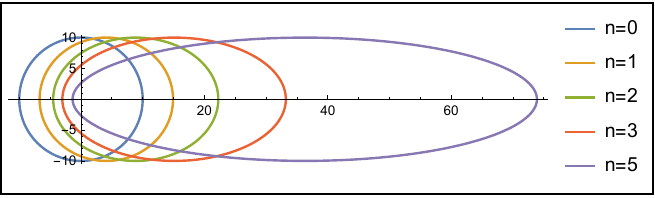}
    \caption{Angular dependence of the  position of event horizon as a function of the stroboscopic time, $n$. We have set:  $ r_+=10\,\, (\beta_{\rm th} = (2\pi)/r_+), \,\, \theta=0.1 \,\, \text{and} \,\, q=1$. The horizon is approaching a point on the boundary. The horizontal axis is the radial coordinate and the angular direction is the angular coordinate of the CFT.}
    \label{fig:your_plot1}
\end{figure}

While it is possible to analytically explore the horizon formula even further, it is more illuminating to have pictorial representations of it. We have presented several plots towards this. For example, in figure \ref{fig:your_plot1}, we have shown the evolution of the Horizon with stroboscopic time using, for the simplest case, $q=1$. In this case, initial $U(1)$ symmetric horizon gets stretched as time progresses, breaks the $U(1) \to {\mathbb Z}_2$ and the resulting peak grows unbounded towards the conformal boundary. The point at which the horizon touches the conformal boundary can be seen as corresponding to the peaks of the CFT stress-tensor expectation value. 
\begin{figure}[h!tbp]  
    \centering
     \begin{minipage}[t]{0.48\textwidth}
     \centering
    \includegraphics[width=\textwidth]{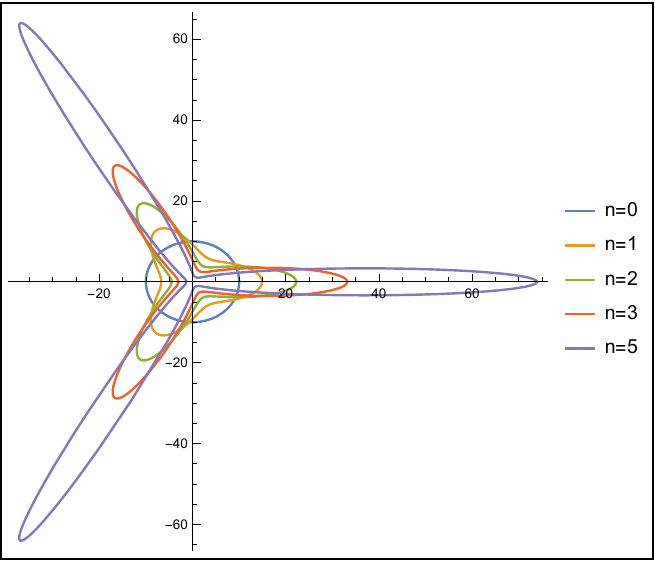}
    \caption{Angular dependence of the  position of event horizon as a function of the stroboscopic time for odd value of $q$. We have set: $ r_+=10 \,\,(\beta_{\rm th} = (2\pi)/r_+), \,\, \theta=0.1 \,\, \text{and} \,\, q=3$. We can see that the number of peaks has increased from the previous case as we increase the value of $q$. The horizontal axis is the radial coordinate and the angular direction is the angular coordinate of the CFT.}
    \label{fig:your_plot2}
    \end{minipage}\hfill
\begin{minipage}[t]{0.48\textwidth}
    \centering
    \includegraphics[width=\linewidth]{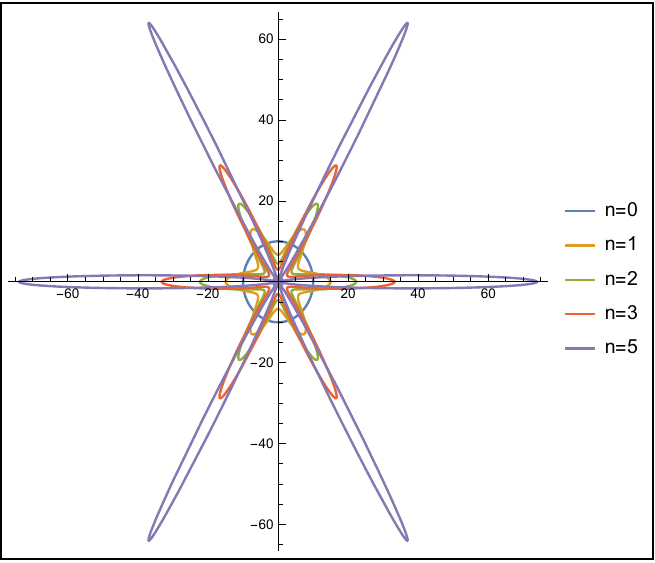} 
        \caption{Angular dependence of the  position of event horizon as a function of the stroboscopic time for even value of $q$. We have set: $ r_+=10 \,\,(\beta_{\rm th} = (2\pi)/r_+), \,\, \theta=0.1 \,\,  \,\, q=6$. Here we can also see that the number of peaks is essentially the same as the value of $q$. The horizontal axis is the radial coordinate and the angular direction is the angular coordinate of the CFT.}
        \label{fig:your_plot3}
    \end{minipage}\hfill
     \end{figure}
\begin{figure}[h!tbp]  
    \centering
     \begin{minipage}[t]{0.48\textwidth}
     \centering
    \includegraphics[width=\textwidth]{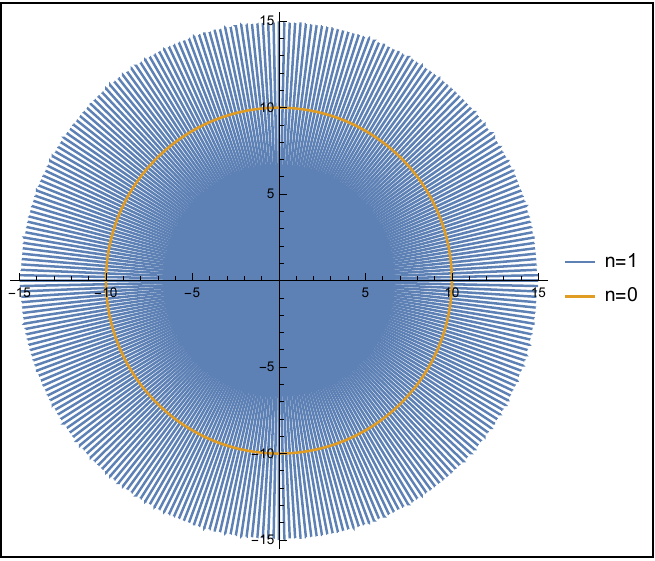}
    \caption{Angular dependence of the  position of event horizon as a function of the stroboscopic time for large value of $q$ in the heating phase. We have set: $ r_+=10 \,\,(\beta_{\rm th} = (2\pi)/r_+), \,\, \theta=0.1 \,\, \text{and} \,\, q=300$. We can see that for large value of $q$ there is a restoration of $U(1)$ symmetry. The horizontal axis is the radial coordinate and the angular direction is the angular coordinate of the CFT.}
    \label{fig:your_plot2_q300}
    \end{minipage}\hfill
\begin{minipage}[t]{0.48\textwidth}
    \centering
    \includegraphics[width=\linewidth]{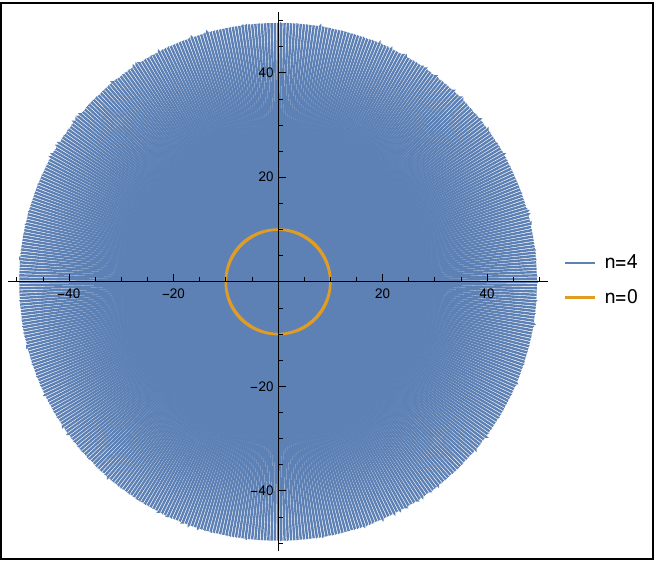} 
        \caption{Angular dependence of the  position of event horizon as a function of the stroboscopic time for large value of $q$ in the heating phase. We have set: $ r_+=10\,\,(\beta_{\rm th} = (2\pi)/r_+), \,\, \theta=0.1 \,\,  \,\, q=500$. Here we can also see that for large value of $q$ there is a restoration of $U(1)$ symmetry. The horizontal axis is the radial coordinate and the angular direction is the angular coordinate of the CFT.}
        \label{fig:your_plot3_q500}
    \end{minipage}\hfill
     \end{figure}

It is further evident from figures \ref{fig:your_plot2} and \ref{fig:your_plot3} that the symmetry breaking pattern of the event horizon is given by $U(1) \to {\mathbb Z}_q$. In the quench-limit, similar symmetry breaking was also studied in \cite{miyata2024hawkingpageentanglementphasetransition, goto2021nonequilibratingblackholeinhomogeneous} This symmetry breaking is explicit, since the driving Hamiltonian itself breaks this symmetry. Nonetheless, it is intriguing to note that in the limit $q\gg 1$, the density of peaks in the event horizon also becomes large. In an effective sense, therefore, in the $q\to \infty$ limit, the ${\mathbb Z}_q$ symmetry again approximates an emergent $U(1)$ symmetry, as is evidenced in figure \ref{fig:your_plot2_q300}, \ref{fig:your_plot3_q500}.\footnote{Technically, this can be traced back to the behaviour of the $\cos\left(\frac{2\pi q x}{L} \right)$ term, in the $q\to \infty$ limit. In this limit the only sensible quantity is : $\int \cos\left(\frac{2\pi q x}{L} \right) dx$, evaluated within a range that is large compared to the variation scale of this function. Such an integral yields a vanishing expectation value. Therefore, in all corresponding observables, one observes an emergent $U(1)$ symmetry. } The universal feature in all these cases is simply that the peaks grow unbounded till it touches the conformal boundary of AdS$_3$ and these points correspond to the peaks in the CFT stress-tensor expectation value.

\subsection{Phase Transition}

In this phase we can also write the explicit metric by putting the expressions of $ \mathscr{L}^{\pm}(x^{\pm})$ in \eqref{eq:2.2}.
\begin{align}
     \mathscr{L}^{\pm}(x^{\pm})=-\frac{q^2\pi^2 }{L^2}+\left(\frac{q^2\pi^2 }{L^2}+\frac{\pi^2 }{\beta^2}\right) \Big[1-2n\tanh{2\theta}\Big[\cos{\frac{2\pi  x^{\pm}}{l}}+\notag \\ n \left(\sin{\frac{2\pi  x^{\pm}}{l}}-1 \right)\tanh{2\theta}\Big]\Big]^{-2} \ . \label{eq:5.8}
\end{align}

In a similar manner, we can calculate the diffeomorphism functions,
$f_{n,\pm}(x^{\pm})$, with the appropriate transformation parameters $a_n$\, and  $b_n$ from section \ref{phasetran_line}, which yields:
\begin{equation}
   f_n(x^-)= \frac{L \left(\tan ^{-1}\left(\frac{1}{\frac{1}{\tan \left(\frac{\pi  q x^-}{L}\right)-1}+n \tanh (2 \theta )}+1\right)-\tan ^{-1}\left(\frac{1}{\frac{1}{\cot \left(\frac{\pi  q x^-}{L}\right)-1}-n \tanh (2 \theta )}+1\right)\right)}{2 \pi  q} \ .
   \label{eq:5.9}
\end{equation}
Thus the horizon location, using \eqref{eq:5.6}, is given by
\begin{equation}
   r(x^-)= \frac{\pi  \sqrt{1-\frac{\beta_{\rm th} ^2 n^2 q^2 \tanh ^2(2 \theta ) (\sin (q x^-)-n \tanh (2 \theta ) \cos (q x^-))^2}{\pi ^2}}}{\beta_{\rm th} -2 \beta_{\rm th}  n \tanh (2 \theta ) (n \tanh (2 \theta ) (\sin (q x^-)-1)+\cos (q x^-))}
   \label{eq:5.10}
\end{equation}
Here we also have set $L=2\pi$.

Let us now provide illustrative evidence of the features of the horizon at the phase transition, using a polar plot as before.
\begin{figure}[h!tbp]  
    \centering
    
    \includegraphics[width=0.8\textwidth]{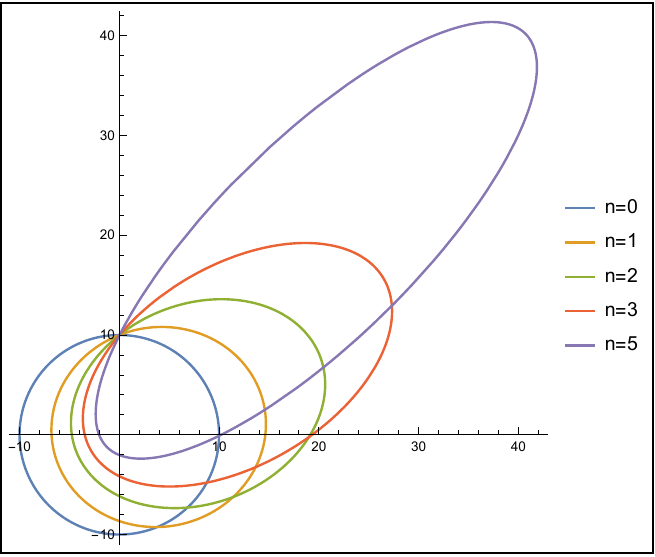}
    \caption{Angular dependence of the  position of the event horizon as a function of the stroboscopic time, $n$. We have set: $ r_+=10\,\, (\beta_{\rm th} = (2\pi)/r_+), \,\, \theta=0.1 \,\, \text{and} \,\, q=1$. With each drive cycles the peak shifts and rotates compared to its starting point. The horizontal axis is the radial coordinate and the angular direction is the angular coordinate of the CFT.}
    \label{fig:your_plot4}
\end{figure}
\begin{figure}[h!tbp]  
    \centering
    
    \includegraphics[width=0.59\textwidth]{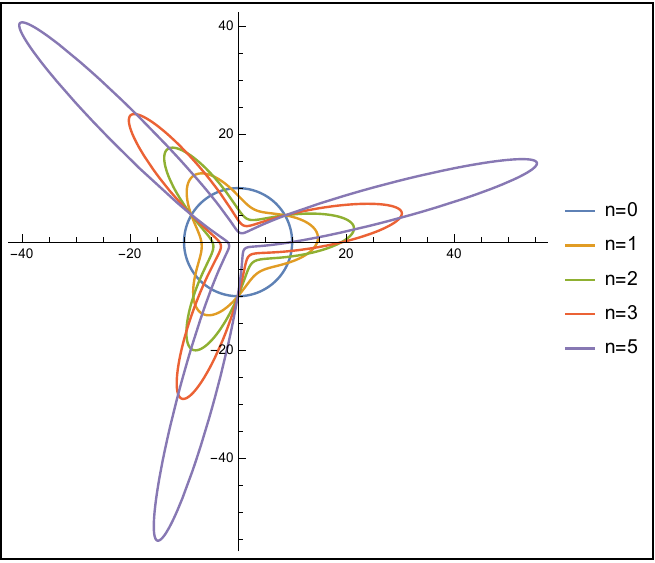}
    \caption{Similar angular dependence of the position of the event horizon as a function of the stroboscopic time for odd value of $q$. We have set: $ r_+=10\,\,(\beta_{\rm th} = (2\pi)/r_+), \,\, \theta=0.1 \,\, \text{and} \,\, q=3$. We can see that the number of peaks has been increased from the previous case as we have increased the value of $q$. The horizontal axis is the radial coordinate and the angular direction is the angular coordinate of the CFT.}
    \label{fig:your_plot5}
\end{figure}
\begin{figure}[ht]
    \centering
    \begin{minipage}[t]{0.48\textwidth}
        \centering
        \includegraphics[width=\textwidth]{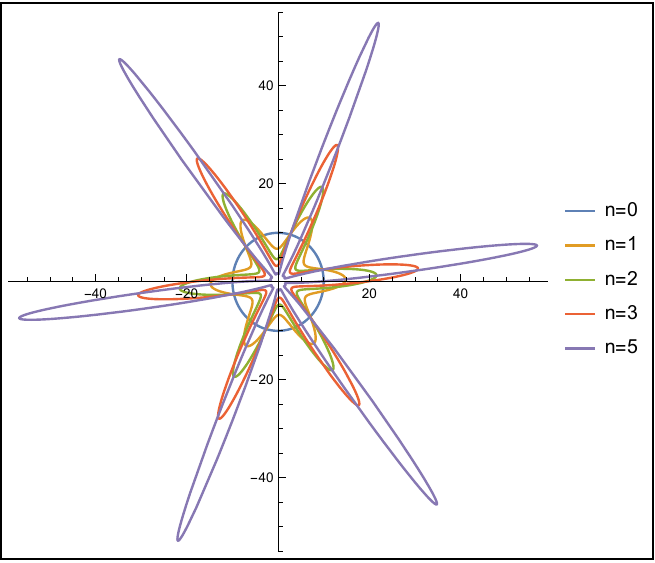} 
        \caption{Angular dependence of the  position of the event horizon as a function of the stroboscopic time for even value of $q$. We have set: $ r_+=10\,\,(\beta_{\rm th} = (2\pi)/r_+), \,\, \theta=0.1 \,\,  \,\, q=6$. There are now six peaks. The horizontal axis is the radial coordinate and the angular direction is the angular coordinate of the CFT.}
        \label{fig:your_plot6}
    \end{minipage}\hfill
    \begin{minipage}[t]{0.48\textwidth}
        \centering
        \includegraphics[width=\textwidth]{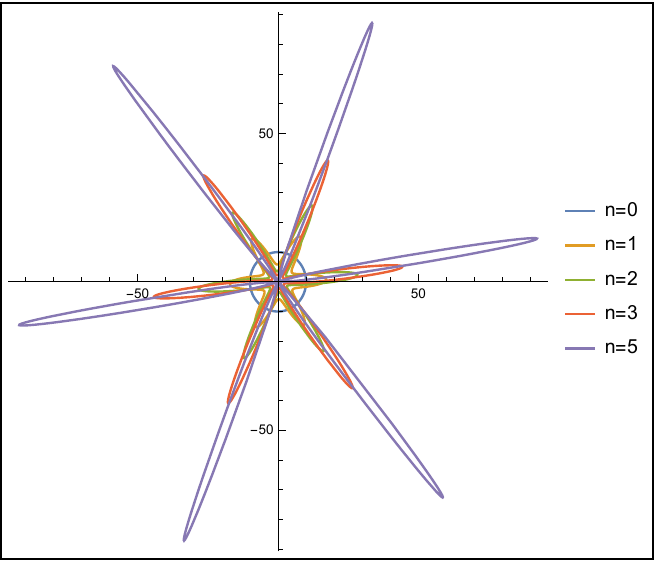} 
        \caption{Angular dependence of the position of the event horizon as a function of the stroboscopic time. We have set: $ r_+=10 \,\,(\beta_{\rm th} = (2\pi)/r_+), \,\, \theta=0.13 \,\, \,\, q=6$. Here, we have a different value of $\theta$ as compared to the other plot. Increasing the $\theta$-value shrinks the width of the peaks.}
        \label{fig:your_plot7}
    \end{minipage}
     
\end{figure}
\FloatBarrier
Our main observations are summarized pictorially in figures \ref{fig:your_plot4},\ref{fig:your_plot5},\ref{fig:your_plot6},\ref{fig:your_plot7}. On the phase transition line also, we observe a symmetry breaking pattern: $U(1) \to {\mathbb Z}_q$, which in the $q\to \infty$ limit should again yield an emergent $U(1)$ symmetric description. Unlike the heating phase, here the peaks shift monotonically in the radial direction (as $n$ increases) and it also rotates along the angular direction. The corresponding parabolic conjugacy class, therefore, has a shift and rotation action on the horizon. The amount of shift depends on the parameter $\theta$ and grows with increasing values of this parameter. Note that, naively, on the phase transition line as well, the peaks grow unbounded in the stroboscopic time. We therefore need a better qualifier to understand the dynamical differences between these two phases.

\begin{figure}[h!tbp]  
    \centering
    
    \includegraphics[width=\linewidth]{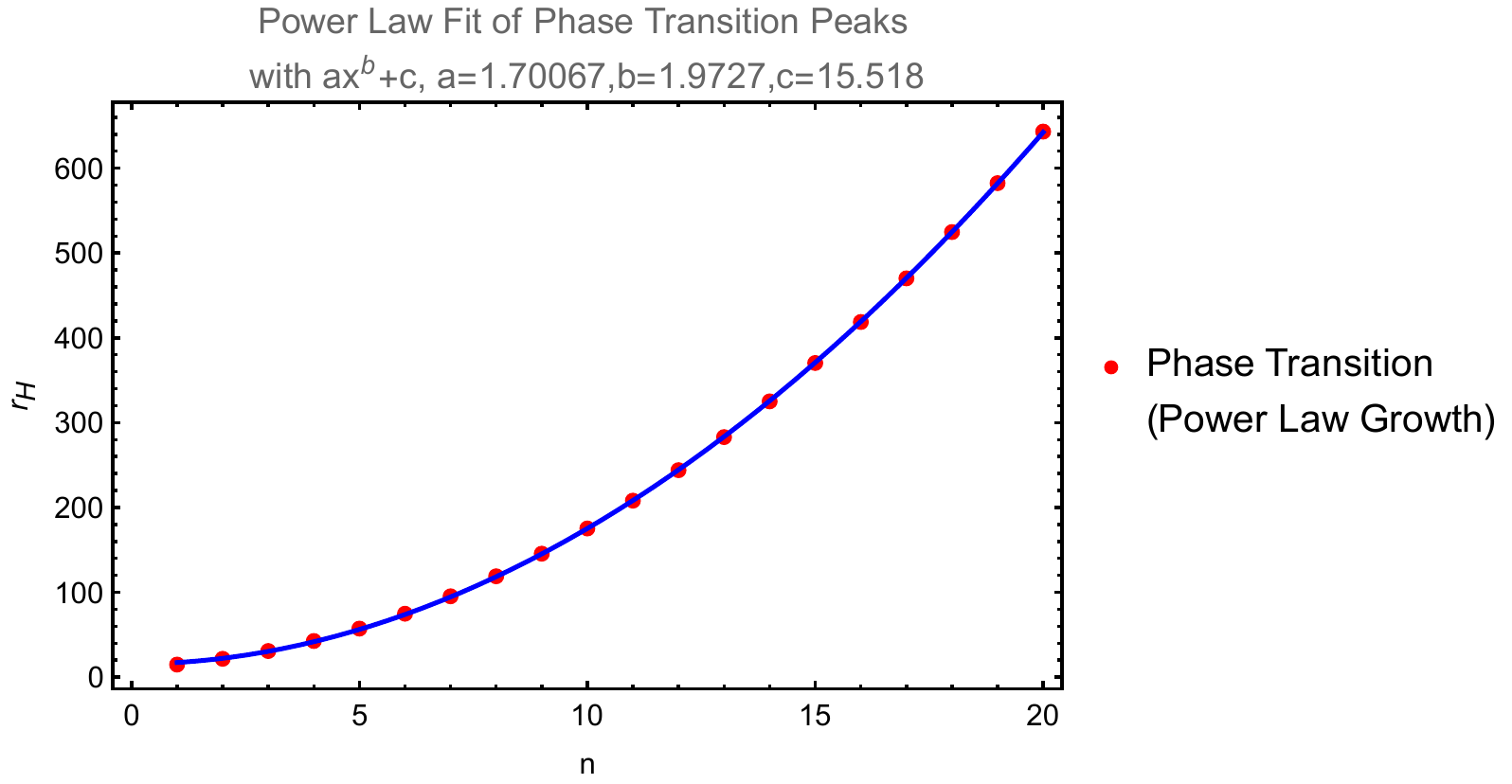}
    \caption{Variations of the Event Horizon Peaks with number of drive cycles. We have set: $ r_+=1.0, \,\, \theta=0.1 \,\, q=1 $.}
    \label{fig:your_plot8}
\end{figure}

\begin{figure}
    \centering
    
    \includegraphics[width=\linewidth]{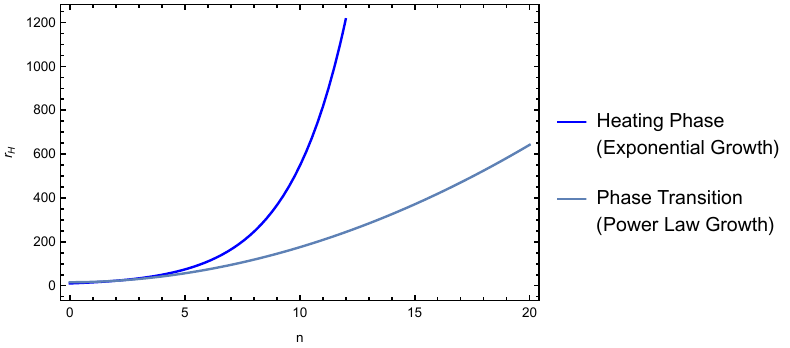}
    \caption{Variations of the Event Horizon Peaks with number of drive cycles. We have set: $ r_+=10, \,\, \theta=0.1 \,\, q=1 $. The peaks of the event horizon in heating phase grows exponentially and shows a power law growth on the phase transition line.}
    \label{fig:your_plot9}
\end{figure}

It is straightforward to check how fast the peak grows as a function of $n$, in the heating phase and on the phase transition line and provide a quantitative comparison between them. This is provided in figure \ref{fig:your_plot8}, in which the growth rate matches with a power-law on the phase transition line. On the other hand, figure \ref{fig:your_plot9} provides a direct comparison between the heating phase and the phase transition line. In the former, the growth is exponential and therefore reaches the boundary that much faster. 

\FloatBarrier

\subsection{Non-Heating Phase}

Let us finally discuss the non-heating phase. This is perhaps the most intriguing scenario, when despite the presence of an event horizon in the initial state, an energy injection will not be absorbed into the horizon. Rather, for the non-heating physics to hold, as the drive progresses, the previously injected energy should instead come out from the event horizon and the event horizon is expected to relax to its initial state.

\begin{figure}[h]
    \centering
    \begin{minipage}[t]{0.48\textwidth}
        \centering
        \includegraphics[width=\textwidth]{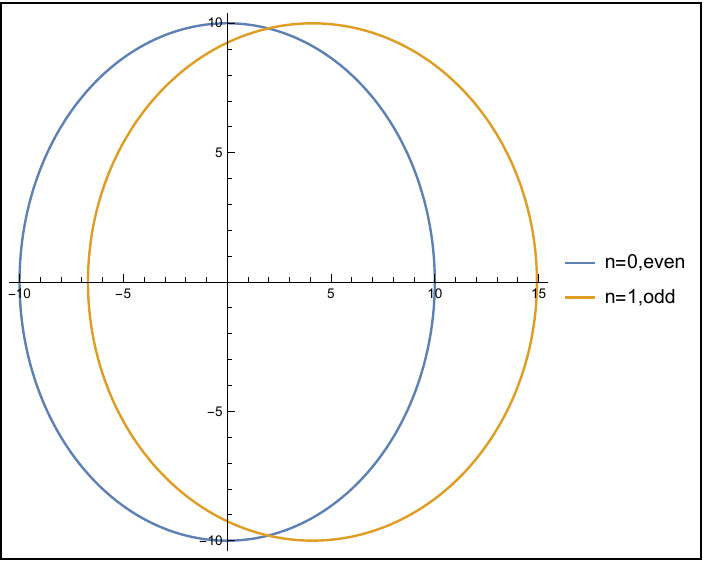} 
        \caption{Angular dependence of the  position of Event Horizon for different number of drive cycles in the bulk CFT with  $ r_+=10\,\,(\beta_{\rm th} = (2\pi)/r_+), \,\, \theta=0.1 \,\,  \,\, q=1 $. The Event horizon Oscillates between two values with number of drives. The horizontal axis is the radial coordinate and the angular direction is the angular coordinate of the CFT. }
        \label{cycles.fig:your_plot10}
    \end{minipage}\hfill
    \begin{minipage}[t]{0.48\textwidth}
        \centering
        \includegraphics[width=\textwidth]{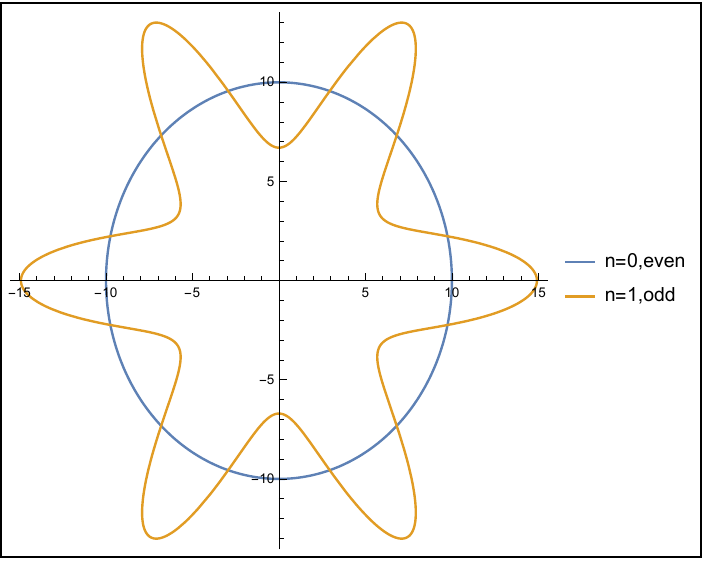} 
        \caption{Angular dependence of the  position of Event Horizon for different number of drive cycles in the bulk CFT with $ r_+=10\,\,(\beta_{\rm th} = (2\pi)/r_+), \,\, \theta=0.1 \,\, \,\, q=6$. The Event horizon Oscillates between two values with number of drive. The number of peaks has increased also as we have increased the value of $q$ (extending the $sl(2,\mathbb{R})$ algebra) similarly to the heating phase. The number of peaks is essentially the same as the value of $q$. The horizontal axis is the radial coordinate and the angular direction is the angular coordinate of the CFT. }
        \label{fig:your_plot11}
    \end{minipage}
     
\end{figure}

Following the same procedure that we have already explained, we plot the behaviour of the event horizon in figures \ref{cycles.fig:your_plot10} and \ref{fig:your_plot11}. While the former plot is for $q=1$, the latter is for $q=5$. While the pattern of the symmetry breaking $U(1) \to {\mathbb Z}_q$ persists here as well, the stroboscopic dynamics of the event horizon is qualitatively different. 
In particular, there is no unbounded growth of the event horizon. Rather, the horizon (the center of the horizon or the peak) shifts along the radial direction under the first part of the non-trivial evolution. Thus, instead of a periodic drive, if we consider a quench limit, then the horizon will radially shift ({\it e.g.}~like the off-center ellipse in figure \ref{cycles.fig:your_plot10}) and will stay there forever. Thus, in the quench limit, the energy that is injected into the system is completely absorbed by the event horizon and the corresponding geometry never relaxes back to its original state. That is why there is no non-heating phase in quench.

For a periodic drive, another drive cycle begins. In this case, the event horizon again slides back to its its original shape and position ({\it e.g.}~in the position of the blue circle in figure \ref{cycles.fig:your_plot10}) releasing the energy that was injected into the system. The time-period of this revival is determined by the time-period of the drive. Note that, the elliptic conjugacy class transformation therefore corresponds to stretching the horizon realizing a symmetry breaking of $U(1) \to {\mathbb Z}_q$ as well as shifting it in one cycle and then applying an inverse transformation in the next cycle. This is technically reflected in the fact that $(M_0 M_1)^{\rm even} \sim 1$ in this case. 

\section{Integral Curves and Fixed Points in the Bulk}

In this section, we will summarize certain salient features of the fixed points of the integral curves. In this case, we have not been able to solve for the integral curves in a closed analytic form, however, the fixed points can be explored in detail. However, for a general $q$, the algebraic equations are still not soluble in full generality. We will set $\alpha=0$, in which limit the structure of the fixed points substantially simplify. Also note that in the $\alpha=0$ limit, only the heating phase is available. We will relegate the details of the integral curves in appendix \ref{appenF} and we will here discuss only the features here. Note that, the fixed points correspond to a bulk observer with a vanishing acceleration. 

\subsection{\texorpdfstring{Fixed Points as a Function of $\textbf{q}$}{}}

The solutions of these equations for $q=1$ are already obtained in \cite{das2024notesheatingphasedynamics}. Here we will discuss what happens for $q>1$ . Recall that for $q=1$ the general solutions are:
\begin{equation}
    \tau=-\frac{\alpha}{2\beta} \ , 
    \label{eq:6.1}
\end{equation}
\begin{equation}
    x^2+z^2=-\frac{\alpha^2-4\beta\gamma}{4\beta^2} \ . 
    \label{eq:6.2}
\end{equation}
When we set $\alpha=0$ and $\gamma=\beta$, we clearly see $\tau=0$ and $z^2+x^2=1$.  
For $q>1$ we will mainly focus on the $\alpha=0$ slice as well as $\gamma=\beta$, which simplifies these equations substantially. Now the equations \eqref{eq:F.2},\eqref{eq:F.3},\eqref{eq:F.4} become:
\begin{align}
  &\hspace*{15mm}  \frac{dz(s)}{ds} =\frac{1}{2} \beta  z \left((-i x)^q+(i x)^q\right) \left(x^{-2q}\right) \left((q+1) \left(x^{2q}\right)-q+1\right) \ ,
    \label{eq:6.3}
    \\
  &  \frac{d\tau(s)}{ds} =-\frac{1}{4} i \beta  \left((-i x)^q-(i x)^q\right) x^{-2 q-1} \left(q z^2 \left((q+1) x^{2 q}-q+1\right)+2 x^{2 q+2}-2 x^2\right) \ ,
  \label{eq:6.4}
  \\
  &   \frac{dx(s)}{ds}=\frac{1}{4} \beta  \left((-i x)^q+(i x)^q\right) x^{-2 q-1} \left(-q z^2 \left((q+1) x^{2 q}+q-1\right)+2 x^{2 q+2}+2 x^2\right) \ .
  \label{eq:6.5}
\end{align}
These are the integral curve equations. 

Let us now classify the solutions into two categories: for even $q$ and for odd $q$. A careful look into these equations reveal the following structure of these equations \eqref{eq:F.2},\eqref{eq:F.3},\eqref{eq:F.4}: for even values of $q$, \eqref{eq:F.3} identically vanishes while \eqref{eq:F.2} and \eqref{eq:F.4} yield a coupled algebraic equation for $\{z,x\}$ and can be solved to obtain two specific values of these coordinates. The corresponding fixed point in the bulk is therefore indeed a point in the bulk geometry. On the other hand, for odd values of $q$ only \eqref{eq:F.3} yields an algebraic equation involving $\{z, x\}$, while the other two equations vanish identically. Therefore, we obtain an equation of a curves in terms of $z$ and $x$.

\subsubsection{\texorpdfstring{Solutions for Even $\textbf{q}$}{}}

For even $q$ the equations \eqref{eq:F.2},\eqref{eq:F.4} become:
\begin{align}
    &\hspace*{25mm} \frac{dz(s)}{ds} =  i^q \beta z x^{-q} \left((q+1) x^{2 q}-q+1\right)=0 \ , \label{eq:6.6}
    \\
    & \frac{dx(s)}{ds} =  i^q \beta x^{-q-1} \left(x^{2 q} \left(x^2-\frac{1}{2} q (q+1) z^2\right)+x^2-\frac{1}{2} (q-1) q z^2\right)=0 \ .\label{eq:6.7}
\end{align}
Comparing with $q=1$ case from \eqref{eq:6.2}, we see that for $\alpha=0$ the RHS becomes $\frac{\gamma}{\beta}$ and subsequently becomes unity when we further set $\gamma=\beta$. To generalize this for even values of $q$, let us use an ansatz: $x=A^{\frac{1}{2q}}$ and $z=B^{\frac{1}{2q}}$, where $A$ and $B$ are constants. We plug this ansatz in the equations and get:
\begin{align}
    &  A =\frac{(q-1)}{(q+1)} \quad \longrightarrow x=\left(\frac{q-1}{q+1}\right)^{\frac{1}{2 q}}  \ , \label{eq:6.8}
    \\
 &    B =\frac{2^q (q-1)^{1-q}}{(q+1)^{q+1}} \quad \longrightarrow z=\frac{\sqrt{2} \left(\frac{q-1}{q+1}\right)^{\frac{1}{2 q}}}{\sqrt{q^2-1}} \ .
\label{eq:6.9}
\end{align}

The solutions are constant points on the $z-x$ plane at the $\tau=0$ slice and the points lie on the intersection of the following two curves:
\begin{eqnarray}
&& z^{2q}+x^{2q}=\frac{(q-1) \left(2^q \left(q^2-1\right)^{-q}+1\right)}{q+1} \ , \label{eq:6.10} \\
&& x = m z \ , \quad m = \left(\frac{B}{A} \right)^{1/2q} \ . \label{eq:6.11}
\end{eqnarray}
This curve in \eqref{eq:6.10} is known as Supercircle or Lamé Curve. At larger values of $q$, the fixed point moves closer to the conformal boundary of AdS, since $z \to 1/q$ in this limit.

\subsubsection{\texorpdfstring{Solutions for Odd $\textbf{q}$}{}}

For odd values of $q$ \eqref{eq:F.3} becomes:
\begin{equation}
    \frac{d\tau(s)}{ds}=\frac{1}{2}  i^{q+1}  \beta x^{-q-1} \left(x^{2 q} \left(q (q+1) z^2+2 x^2\right)-(q-1) q z^2-2 x^2\right)=0 \ . \label{eq:6.12}
\end{equation}
Solving this we obtain an equation of a curve:
\begin{equation}
    z=\frac{\sqrt{2} \sqrt{x^2 \left(1-x^{2 q}\right)}}{\sqrt{q \left((q+1) x^{2 q}-q+1\right)}} \ . \label{eq:6.13}
\end{equation}
For $q=1$ this reduces to the circle $z^2+x^2=1$\cite{das2024notesheatingphasedynamics}. As before, also note that as $q \rightarrow\infty$ $z\to 0$.

It is crucial to emphasize a couple of points here. The first assumption we made while solving the equations was that $z \neq 0$, which is reasonable given that we are solving the equations in the bulk.\footnote{Therefore we are explicitly ignoring any fixed point on the $z=0$ slice.}
Furthermore we have chosen $\alpha=0$ for analytical control on the solutions. That means $\delta=-4\beta\gamma \leq 0$ for any values of $\beta$ and $\gamma$. So \eqref{eq:D.5} only has solutions in the Heating phase and, in the limiting case, on the phase boundary. This indicates that a bulk observer can access the entire spacetime rather than only a sub-region such as the entanglement wedge, since there are no bulk fixed points in the non heating phase. This holds for arbitrary values of $q$, generalizing the case of $q=1$ in \cite{das2024notesheatingphasedynamics}.

\subsubsection{Fixed points \& Its Physical Interpretation}

Since the general $sl^{(q)}(2,\mathbb{R})$ valued Floquet Hamiltonian in the heating phase, can be mapped to the modular Hamiltonian of a CFT sub-region, at the boundary the fixed points are always guranteed to lie on the corresponding Ryu-Takayanagi (RT) surface. For $q=1$, the Floquet Hamiltonian extends to an exact Killing vector in the bulk and therefore the fixed point generates the entire curve of the RT-surface. For asymptotic Killing vectors, however, this is not expected since deeper in the bulk the corresponding asymtptotic Killing vectors differ from symmetry generators. As we will now observe, the fixed points associated to the asymptotic Killing vectors will instead generate the RT surface near the conformal boundary and deeper in the bulk will deviate from it. The deviation, physically, corresponds to an actual work done by the approximate symmetry generators deeper in the bulk.

A quantitative measure of this work done can be characterized by evaluating the distance of the exact fixed points for $q>1$ cases from the circle: $x^2+ z^2 =1 $. We calculated this distance along the straight line $z=\sqrt{\frac{2}{q^2-1}}x$, which is given by
\begin{align}
d=\sqrt{\frac{q^2+1}{2}}\ln( \left(\frac{q+1}{q-1}\right)^{\frac{1}{2q}}\sqrt{\frac{q^2-1}{q^2+1}}) \ .  \label{eq:6.14}
\end{align}
\begin{figure}[ht]
    \centering
    \begin{minipage}[t]{0.48\textwidth}
        \centering
        \includegraphics[width=\textwidth]{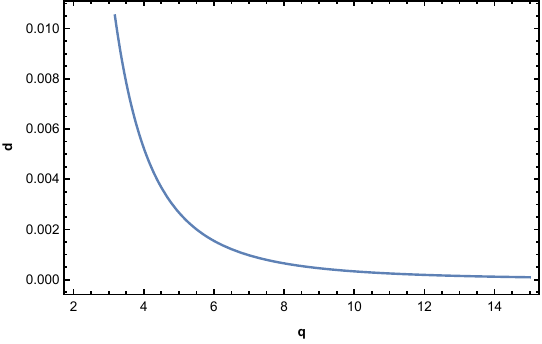} 
        \caption{We plotted this distance \eqref{eq:6.14} with $q$. As $q$ increases we can see the distance becoming zero, implying that the fixed points lies exactly on the semi circle.}
        \label{fig:your_plot20}
    \end{minipage}\hfill
    \begin{minipage}[t]{0.48\textwidth}
        \centering
        \includegraphics[width=\textwidth]{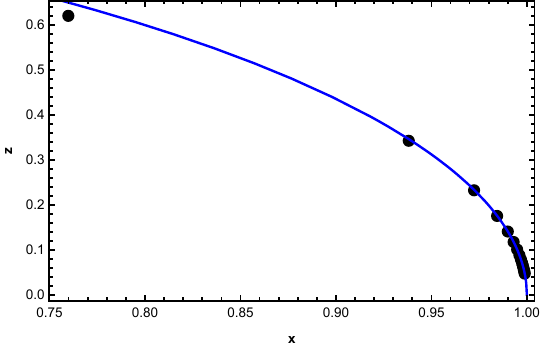} 
        \caption{We have showed the position of the fixed points for even values of $q$ from 2 to 30. For higher values of $q$, the points exactly lies on the semi circle.}
        \label{fig:your_plot21}
    \end{minipage}
     
\end{figure}
The distance decreases with increasing $q$, since the fixed points themselves move closer to the boundary. In this sense, there is a clear notion of bulk reconstruction of the RT-surface in terms of the fixed points corresponding to $q>1$.
\begin{figure}[ht]
\centering
    \begin{minipage}[t]{\textwidth}
        \includegraphics[width=\textwidth]{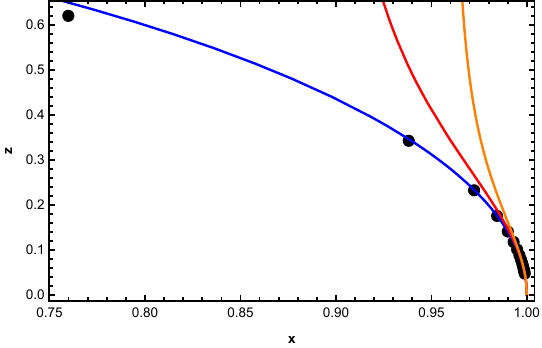} 
        \caption{We plotted the fixed point solutions for all $q$ (even values from $2$ to $30$, odd values $3$ and $5$ and $q=1$). The blue curve is the semicircle, the other curves are for odd $q$ and the dots are for even $q$. We can clearly see near the boundary the fixed points exactly lies on the semi-circle.}
        \label{fig:your_plot22}
    \end{minipage}
    \end{figure}
For odd values of $q$, we obtain a curve, which are pictorially shown together with the even $q$ cases in figure \ref{fig:your_plot22}. It is clear, pictorially, that for odd and the even $q$ values deviate on different directions from the $q=1$ semi-circle. At present, it is not clear to us what this precisely means.

\section{Conclusions}

In this article, we analyzed two complimentary approaches to understand the bulk geometric dual picture of a driven CFT (with a large central charge), when the drive Hamiltonian is constructed from $sl^{(q)}(2,\mathbb{R})$ generators. Our results follow the framework of \cite{deBoer:2023lrd} and \cite{das2022branedetectorsdynamicalphase}. Under such a drive the vacuum also evolves\footnote{Unlike the special case $q=1$, which corresponds to the global part of the Virasoro algebra which keeps the vacuum invariant.} and therefore vacuum correlators display non-trivial dynamical features. This corresponds to inserting physical graviton degrees of freedom in the bulk, albeit only at the conformal boundary. Physically, this corresponds to inserting an energy into the system.

Starting with an initial thermal state, we demonstrated that a general $sl^{(q)}(2,\mathbb{R})$ drive induces a flowery structure to the event horizon, with $q$-petals. The growth rate of these $q$-petal flowery horizons clearly demarcate the phase of the corresponding CFT. Furthermore, corresponding to each conjugacy classes (or the signature of the quadratic Casimir), the evolution of the event horizon correspond to distinct geometric transformations in the bulk. In the heating phase ({\it i.e.}~hyperbolic conjugacy class), the petals grow exponentially in the stroboscopic time towards the conformal boundary. In the non-heating phase ({\it i.e.}~elliptic conjugacy class), the petals of the horizon oscillate as a function of the stroboscopic time. Finally, on the phase transition line ({\it i.e.}~parabolic conjugacy class), the petals of the horizon rotate as well as exhibit a power-law growth in the stroboscopic time. These provide a co-ordinate-dependent but an explicit pictorial realization of the various conjugacy classes of transformations, in the bulk geometry.

Observe that, on the phase transition line, the petals undergo a macroscopic rotation even at early stroboscopic times, as compared to the heating phase. Therefore, by simply looking at the dynamical behaviour of the petals, we can distinguish between the transition line and the heating phase, at a given value of $n$, even if it is not large. This distinction at early stroboscopic times is not manifestly available in the dual CFT. Note that, the distinction between the heating phase and the phase transition line blurs in the $q\to \infty$ limit, in which case the rotation of the horizon becomes invisible since the $U(1)$ symmetry is effectively restored. Nonetheless, the growth of the peaks with respect to the stroboscopic time still retains its distinction.

We have also demonstrated that for a general $q$, the Floquet Hamiltonian in the heating phase can be mapped to a modular Hamiltonian of a sub-region in the CFT. However, the bulk extension of the corresponding Hamiltonian is only an exact symmetry for $q=1$. With this exact symmetry, it is known that the zero acceleration observer which is generated by the corresponding bulk Killing vector, actually generates the  Ryu-Takayanagi surface. For a general $q$, this is not the case, since deeper in the bulk the asymptotic symmetries do not correspond to Killing vectors. The integral curves capture this physics in a rather nice manner. The fixed points for the integral curves essentially approximate the Ryu-Takayanagi surface near the boundary and depart further in the deep IR. We have provided explicit realizations of this, in terms of pictures and curves.

The geometric picture above is already suggestive that there is a non-trivial ``work done" when $q\not =1$. This is expected, since $q\not =1$ corresponds to actual insertions of boundary gravitons and a cost associated with it. This work done can also be viewed as an IR-effect, which is expected to affect the long-time dynamics of the system. Given the explicit curves (odd $q$) or bulk points (even $q$), it may be possible to relate a covariant distance from the corresponding Ryu-Takayanagi surface and relate it to the vacuum expectation value of the stress-tensor. We hope to come back to this issue in future.

The construction above can be seen as an explicit realization of how a bulk observer is generated, given the CFT Hamiltonian. The asymptotic Killing vectors are equipped to describe bulk points that are away from the Ryu-Takayanagi surfaces. Therefore, given a linear sub-region, $q=1$ describes an observer within the entanglement wedge of the sub-region, while higher $q$ generators are capable of describing points outside this wedge. In this sense, the modular Hamiltonian has non-trivial and infinitely many bulk extensions and is capable of capturing an infinitely larger region in the bulk. This, of course, results from the existence of the infinite dimensional asymptotic symmetry operations in the bulk. It will be very interesting to understand this aspect better.

It will also be interesting to understand and explicitly realize the full dynamical geometry that evolves as the CFT is being driven. Here, we have side-stepped this issue by focusing only on the dynamics of the event horizon, which we obtained by a chain of coordinate transformations. We have been agnostic about how {\it e.g.}~causality constraints in the bulk manifest in the evolution of the geometry. It will be very useful to capture this physics in full details, which we leave for future work.

\end{sloppypar}

\section{Acknowledgements}

We would like to thank Diptarka Das, Sumit R.~Das, Krishnendu Sengupta for collaboration in related topics. We would like to specially thank Jani Kastikainen for very useful discussions and comments. We would further like to thank Suchetan Das, Bobby Ezhuthachan, Abhijit Gadde, Gautam Mandal, Shiraz Minwalla, Somnath Porey, Baishali Roy, Suman Das, Sandip Trivedi for several discussions that are relevant for this work. JD would like to thank the Organizers of NSM 2024, held at Indian Institute of Technology Ropar, and DAE-HEP Symposium 2024, held at Banaras Hindu University, Varanasi, where the work has been discussed. AK is partially supported by DAE-BRNS 58/14/12/2021-BRNS and CRG/2021/004539 of Govt. of India. AK further acknowledges the support of Humboldt Research Fellowship for Experienced Researchers by the Alexander von Humboldt Foundation and for the hospitality of the Theoretical Physics III, Department of Physics and Astronomy, University of Wurzburg during the course of this work.

\appendix

\section{CFT on a Curved Manifold}\label{appenA}

In this Appendix, we briefly collect and review basics on how we can view the driven CFT framework in terms of a standard CFT defined on a curved manifold. This has already been discussed in detail in {\it e.g.}~\cite{miyata2024hawkingpageentanglementphasetransition,Bai:2024azk}. To begin with, we want to consider the following Hamiltonian:
\begin{eqnarray}
    H_\theta = \int_0^L dx \left[1 - \tanh(2\theta) \cos\left( q\frac{2\pi x}{L}\right) \right]  T_{00}(x) \ , \label{eq:A.1}
\end{eqnarray}
which can be recast as:
\begin{eqnarray}
    && H_\theta = \int_0^L dx \sqrt{- {\rm det} \gamma} T_{00}(x) \ , \label{eq:A.2} \\
    && ds^2 = \gamma_{ab} d\zeta^a d\zeta^b = - f(\theta, x)^2 dt^2 + dx^2 \ , \label{eq:A.3} \\
    && f(\theta, x) =  \left[1 - \tanh(2\theta) \cos\left( q\frac{2\pi x}{L}\right) \right] \ . \label{eq:A.4}
\end{eqnarray}
The two-dimensional curved manifold depends on $\theta$, which is the drive parameter. The curvature of this manifold is simply given by $R_2 = - 2 \partial_x^2 f(x,\theta) / f(x, \theta)$. Given the functional behaviour $f(x,\theta)$, it can be easily checked that the curvature is an oscillating function of $x$. In the $q\to \infty$ limit, the number of oscillations diverge, and therefore the curvature is given by suitably averaging over a range of the position variable. This yields a constant, negative curvature: 
\begin{eqnarray}
    \frac{1}{\Delta x} \int_x^{x+\Delta x} R_2 dx = - \frac{1}{\Delta x} \int_x^{x+\Delta x} \frac{2 \partial_x^2 f(x,\theta)}  {f(x, \theta)} dx< 0 \ , \label{eq:A.5}
\end{eqnarray}
where $\Delta x$ is a width which is larger compared to the variation-scale of the oscillation, $1/q$. Interestingly, at large $q$ the CFT can be viewed to be defined on an AdS$_2$ manifold. This has a similarity to a doubly Holographic system, if the CFT has a large central charge.

This will hold for generic values of the drive parameter $\theta$, except at $\theta\to \infty$, in which case, the curvature is unbounded from below. On the other hand, at $\theta =0$, the curvature vanishes identically, and therefore we recover that the CFT is defined on a flat manifold, as expected. Varying $q$, can be viewed as lowering the averaged Ricci-curvature of the manifold, in general.

\section{\texorpdfstring{\boldmath $sl^{(q)}(2,\mathbb{R})$ Locally as $sl(2,\mathbb{R})$}{}}\label{appenB}

Recall that the Virasoro algebra is given by
\begin{eqnarray}
    \left[L_m, L_n \right] = (m-n)L_{m+n} + \frac{c}{12} m(m^2-1) \delta_{m+n} \ , \label{eq:B.1}
\end{eqnarray}
from which it is easy to recover the global $sl(2,\mathbb{R})$ algebra with elements $\{L_0, L_{\pm 1}\}$:
\begin{eqnarray}
\left[L_1, L_0 \right] = L_1 \ , \quad \left[L_0, L_{-1} \right] = L_{-1} \ , \quad \left[L_1, L_{-1} \right] = 2 L_0 \ . \label{eq:B.2}
\end{eqnarray}
Note that, at the algebraic level, $\{L_0, L_{\pm q}\}$ also forms a closed sub-algebra:
\begin{eqnarray}
    \left[L_q, L_0 \right] = q L_q \ , \quad \left[L_0, L_{-q} \right] = q L_{-q} \ , \quad \left[L_q, L_{-q} \right] = 2 q L_0 + \frac{c}{12} q(q^2-1)\ . \label{eq:B.3}
\end{eqnarray}
This is the $sl^{(q)}(2,\mathbb{R})$ sub-algebra. With the following re-definition, we can {\it locally} express this sub-algebra as an $sl(2,\mathbb{R})$ algebra:
\begin{eqnarray}
    {\mathcal L}_0 = \frac{1}{q} \left(L_0 + \frac{c}{24} (q^2-1) \right)  \ , \quad {\mathcal L}_{\pm q} = \frac{1}{q} L_{\pm q} \ . \label{eq:B.4}
\end{eqnarray}
It is easy to check that $\{{\mathcal L}_0, {\mathcal L}_{\pm q}\}$ satisfies an $sl(2,\mathbb{R})$ algebra\cite{Jiang:2024hgt,Caputa:2022zsr}. More precisely $sl^{(q)}(2,\mathbb{R})$ is isomorphic to a $q$-fold cover of $sl(2,\mathbb{R})$, which is also reflected in the corresponding map from a cylinder to the $q$-sheeted Riemann surface: $z= {\rm exp}\left( \frac{2\pi i q w}{L} \right)$.

\section{Detailed Calculations of the Event Horizon}\label{appenC}

In this appendix, we summarize the chain of coordinate transformations and associated algebraic computations that determine the event horizon and its dynamical behaviour.

The AdS$_3$ Rindler horizon is given by $X^2+Z^2=1\,\, , \,\, T=0$ in Poincaré AdS$_3$ coordinates. So we can write $X^{+}=X^{-}=X$. As we are interested in understanding the evolution of event horizon at $t=$constant slices we obtain: $F^+(x^+)=F^-(x^-)=F(x)$. So the \eqref{eq:2.6} becomes:
\begin{equation}
    X=F(x)-\frac{2F^{\prime}(x)^2 F^{\prime\prime}(x)}{4r^2 F^{\prime}(x)^2+F^{\prime\prime}(x)^2 } \ , \quad  
  \,\,Z=\frac{4r F^{\prime}(x)^3}{4r^2 F^{\prime}(x)^2+F^{\prime\prime}(x)^2 } \ .  \label{eq:C.1}
\end{equation}
Now plugging these expression in $X^2+Z^2=1$ we get:
\begin{equation}
    \frac{4 \left(F^{\prime}(x)^4-F(x) F^{\prime}(x)^2 F^{\prime\prime}(x)\right)}{F^{\prime\prime}(x)^2+4 r^2 F^{\prime}(x)^2}+F(x)^2=1 \ . \label{eq:C.2}
\end{equation}
Solving for $r$ we now have:
\begin{equation}
    r_{\rm H}=\frac{\sqrt{4 F(x) F^{\prime}(x)^2 F^{\prime\prime}(x)-4 F^{\prime}(x)^4-\left(F(x)^2-1\right) F^{\prime\prime}(x)^2}}{2 \sqrt{\left(F(x)^2-1\right) F^{\prime}(x)^2}} \ .  \label{eq:C.3}
\end{equation}
Now \eqref{eq:5.2} gives us the functional expression of $F(x)$ for our case. Using these we calculate:
\begin{align*}
& \hspace*{25mm} F^{\prime}(x)=\frac{\pi}{\beta_{\rm th}}  f_n^{\prime}(x) \sech^2{\left(\frac{\pi  f_n(x)}{\beta_{\rm th} }\right)} \ , 
\\ 
& F^{\prime\prime}(x)=\frac{\pi \, \text{sech}^2\left(\frac{\pi  f_n(x)}{\beta_{\rm th} }\right) \left(\beta_{\rm th}  f_n''(x)-2 \pi  f_n'(x){}^2 \tanh \left(\frac{\pi  f_n(x)}{\beta_{\rm th} }\right)\right)}{\beta_{\rm th} ^2} \ .
\end{align*}
Using these, we get the final expression of the event horizon \eqref{eq:5.7}.

We now demonstrate that starting from a vacuum state, rather than a thermal state, leads to no evolution of the horizon. This is because the bulk dual of vacuum states in 2D CFT corresponds to the AdS$_3$ Poincaré patch, which does not possess a horizon.

From \eqref{eq:5.2} we have for vacuum state:
\begin{align}
& F(x)=\tanh{\left(\sqrt{\frac{12\pi \langle T \rangle}{c}}f_{n,(x)}\right)} \ ,  \label{eq:C.4}
   \\
   & \implies F(x)= i \tan{\left(\frac{\pi f(x)}{L}\right)} \ . \label{eq:C.5}
\end{align}
Here, we have used the stress-tensor expectation value corresponding to the vacuum state $\langle T \rangle=-\frac{\pi^2 c}{6L^2}$. Now, if we use the \eqref{eq:5.7}, we get:
\begin{equation}
    r(x)=\frac{\sqrt{-L^2 f''(x)^2-4 \pi ^2 f'(x)^4}}{2 L f'(x)} \ . \label{eq:C.6}
\end{equation}
Since the numerator involves the square root of a negative quantity, there are no real values for the horizon, making it impossible to track its evolution with stroboscopic time. Therefore, we start with a thermal state instead of a vacuum state. In the bulk dual of a thermal state, a black hole is already present, allowing us to effectively track the evolution of the horizon with stroboscopic time.

\section{Heating Phase Floquet Hamiltonian as a Modular Hamiltonian}\label{appenD}

Note, in \cite{das2024notesheatingphasedynamics}, the heating phase Floquet Hamiltonian has been already mapped to the Floquet Hamiltonian corresponding to a subregion $(R_1 , R_2)$ in the vacuum states in a CFT which is defined on a cylinder of circumference $L$. The upshot is that the Floquet evolution can then be identified with the associated modular evolution. Since a CFT thermal state is related to the vacuum state by a simple conformal transformation, it should be also possible to draw a similar conclusion, when the initial CFT state is taken to be the thermal state, which is then subjected to the Floquet evolution. 

Towards that, recall that the modular Hamiltonian\footnote{Note that, in a precise sense, this is the extended modular Hamiltonian\cite{das2024notesheatingphasedynamics}.} for a subregion $(R_1 , R_2)$ is given by\cite{Cardy:2016fqc}
\begin{equation}
    K=\frac{L}{\pi}\int_{0}^{L}\frac{\sin{\frac{\pi(x-R_1)}{L}}\sin{\frac{\pi(R_2-x)}{L}}}{\sin{\frac{\pi(R_2-R_1)}{L}}}T_{00}(x)dx \label{eq:D.1} \ .
\end{equation}
This can be rewritten as:
\begin{align}
     K=\frac{L}{2\pi}\frac{1}{\sin{\frac{\pi(R_2-R_1)}{L}}}\int_{0}^{L} \Bigg[\cos{\frac{\pi(R_1+R_2)}{L}}\cos{\frac{2\pi x}{L}}+\sin{\frac{\pi(R_1+R_2)}{L}}\sin{\frac{2\pi x}{L}} \notag \\ -\cos{\frac{\pi(R_2-R_1)}{L}} \Bigg]T_{00}(x)dx \ .
     \label{eq:D.2}
\end{align}
Here, we considered the Floquet Hamiltonian in terms of $sl^{(q)}(2,R)$ generators:
\begin{equation}
    H=(\alpha L_0+\beta L_q+\gamma L_{-q})+\text{c.c}-\frac{c}{12} \ .
    \label{eq:D.3}
\end{equation}
Let us now use the  definition of the Virasoro generators:
$$
L_q=\frac{c}{24}\delta_{q,0}+\frac{L^{\prime}}{2\pi}\int_0^{L^{\prime}}
\frac{dx}{2\pi}\exp{\frac{2\pi iqx}{L^{\prime}}}T(x) \ ,
$$
$$
\bar{L}_q=\frac{c}{24}\delta_{q,0}+\frac{L^{\prime}}{2\pi}\int_0^{L^{\prime}}
\frac{dx}{2\pi}\exp{-\frac{2\pi iqx}{L^{\prime}}}\bar{T}(x) \ , 
$$
with $T(x)=\pi(T_{00}(x)+T_{01}(x))$ and $\bar{T}(x)=\pi(T_{00}(x)-T_{01}(x))$.

To compare \eqref{eq:D.2} with \eqref{eq:D.3} we set $L^{\prime}=qL$ and use the relation that 
\begin{eqnarray}
\int_0^{qL}\frac{dx}{2\pi}\exp{-\frac{2\pi ix}{L}}T(x)
=q\int_0^{L}\frac{dx}{2\pi}\exp{-\frac{2\pi ix}{L}}T(x) \ .  \label{eq:D.4}
\end{eqnarray}
This is due to the fact $q\in \mathbb{Z}^+$ and because of the periodic boundary condition on stress tensor: $T(x+L)=T(x)$. So we get:
\begin{align}
    \alpha=\frac{1}{q^2}\cot{\frac{\pi(R_1-R_2)}{L}} \ , \notag
    \\
    \beta+\gamma=-\frac{1}{q^2}\frac{\cos{\frac{\pi(R_1+R_2)}{L}}}{\sin{\frac{\pi(R_1-R_2)}{L}}} \ , \notag
    \\
    i(\beta-\gamma)=-\frac{1}{q^2}\frac{\sin{\frac{\pi(R_1+R_2)}{L}}}{\sin{\frac{\pi(R_1-R_2)}{L}}} \ .
    \label{eq:D.5}
\end{align}
Rewriting: $\alpha=a\,,\, \beta+\gamma=b\,,\, i(\beta-\gamma)=c $ then the relation $\alpha^2-4\beta\gamma$ becomes $a^2-b^2-c^2$. Calculating this, we obtain: $\alpha^2-4\beta\gamma=a^2-b^2-c^2=-\frac{1}{q^4}<0$.

Since the quadratic Casimir is negative, it immediately follows that the Hamiltonian in the heating phase can be mapped to the extended modular Hamiltonian of an interval in the vacuum state of a CFT, defined on a ring of length $L=\frac{L^\prime}{q}$ where $L^\prime$ is the length associated to the Floquet CFT.

One can now generalize this using the thermal state as the initial state. We consider the modular Hamiltonian for a subregion $(R_1 , R_2)$ for the thermal states:
\begin{equation}
    K_{\text{thermal}}=\frac{\beta_{\rm th}}{\pi}\int_{0}^{L}\frac{\sinh{\frac{\pi(x-R_1)}{\beta_{\rm th}}}\sinh{\frac{\pi(R_2-x)}{\beta_{\rm th}}}}{\sinh{\frac{\pi(R_2-R_1)}{\beta_{\rm th}}}}T_{00}(x)dx \ . \label{eq:D.6}
\end{equation}
To obtain equivalent relations to \eqref{eq:D.5}, we replace $\frac{\pi}{L}$ by $\frac{\pi}{\beta_{\rm th}}$\footnote{In this section we use $\beta$ as a parameter in the Floquet Hamiltonian and $\beta_{\rm th}$ denotes the inverse temperature for the thermal state.} and trigonometric functions by hyperbolic functions. This yields:
\begin{align}
    \alpha=\frac{i}{q^2}\coth{\frac{\pi(R_1-R_2)}{\beta_{\rm th}}} \ , \notag
    \\
    \beta+\gamma=-\frac{i}{q^2}\frac{\cosh{\frac{\pi(R_1+R_2)}{\beta_{\rm th}}}}{\sinh{\frac{\pi(R_1-R_2)}{\beta_{\rm th}}}} \ , \notag
    \\
    i(\beta-\gamma)=-\frac{1}{q^2}\frac{\sinh{\frac{\pi(R_1+R_2)}{\beta_{\rm th}}}}{\sinh{\frac{\pi(R_1-R_2)}{\beta_{\rm th}}}} \ . 
    \label{eq:D.7}
\end{align}
Given these, we still obtain: $\alpha^2-4\beta\gamma=-1/q^4 <0$. So, the heating phase Hamiltonian corresponding to initial thermal state maps to the modular Hamiltonian for a subregion $(R_1 , R_2)$ for the thermal states, for arbitrary values of $q$. It is noteworthy that for larger and larger values of $q$, the subregion length $(R_1-R_2)$ decreases if we hold the parameters $\{\alpha, \beta,\gamma\}$ constant. {\it A priori}, the length of the subsystem can be decreased arbitrarily by increasing $q$ to arbitrarily large integers. All the way to a UV-regulator in the theory. In this case, the corresponding entanglement entropy captures the universal UV-divergent contribution and the Floquet Hamiltonian at large enough $q$ can be interpreted to be the associated modular Hamiltonian corresponding to the UV-divergent entanglement entropy.

Note also that, the Floquet evolution is identical to the modular evolution. This modular dynamics can be viewed explicitly by choosing a conformal frame\cite{das2024notesheatingphasedynamics}:
\begin{eqnarray}
&&    \omega = \tau + i x = \frac{\sqrt{|\alpha^2 - 4 \beta \gamma|}}{2\beta} \tan \left(\frac{z}{2} \sqrt{|\alpha^2 - 4 \beta \gamma|}\right) \ ,  \label{eq:D.8} \\
&&     \bar{\omega} = \tau + i x = \frac{\sqrt{|\alpha^2 - 4 \beta \gamma|}}{2\beta} \tan \left(\frac{\bar{z}}{2} \sqrt{|\alpha^2 - 4 \beta \gamma|}\right) \ , \label{eq:D.9}
\end{eqnarray}
where the size of the cylinder is now given by
\begin{eqnarray}
    \ell = 2 \log \left[\frac{L}{q \pi \epsilon} \sin\left( \frac{\pi q}{L} \left( R_1 - R_2\right)\right) \right] \ , \label{eq:D.10}
\end{eqnarray}
where $\epsilon$ is a UV-regulator that removes a small region around the end-points of the sub-region. It is evident that both $\epsilon\to 0$ limit as well as $q\to \infty$ limit sends $|\ell \to \infty|$. The corresponding spectrum therefore is expected to become continuous in both limits. 

\section{\texorpdfstring{Asymptotic Killing Vector Fields in Poincar\'{e} $\text{AdS}_3$}{}}\label{appenE}

In this appendix, we summarize the explicit asymptotic Killing vectors in the AdS$_3$ Poincaré coordinate.

We start with the Euclidean Bañados metric in Poincaré coordinate as in \cite{Anand:2017dav}
\begin{equation}
    ds^2=\frac{dz^2+d\xi d\bar{\xi}}{z^2}-\frac{1}{2}S(f,\xi)d\xi^2-\frac{1}{2}\bar{S}(\bar{f},\xi)d\bar{\xi}^2+\frac{z^2}{4}S(f,\xi)\bar{S}(\bar{f},\bar{\xi})d\xi d\bar{\xi}\ . \label{eq:E.1}
\end{equation}
Where $\xi=\tau+ix$ and $\bar{\xi}=\tau-ix$ and $S,\bar{S}$ are the Schwarzian derivatives of the diffeomorphisms $f$ and $\bar{f}$ as defined in \eqref{eq:2.4}. This diffeomorphisms transform pure AdS$_3$ into asymptotically AdS$_3$ geometries with non zero stress tensors.
\\
Now the asymptotic killing vectors in these coordinates has the form:
\begin{equation}
    L_q=\frac{(q+1)z\xi^q}{2}\partial_z+\frac{\xi^{q-1}((q^2+q+\xi^2 S(\xi))\bar{S}(\bar{\xi})z^4-4\xi^2)}{z^4S(\xi)\bar{S}(\bar{\xi})-4}\partial_\xi+
    \frac{2q(q+1)z^2 \xi^{q-1}}{z^4S(\xi)\bar{S}(\bar{\xi})-4}\partial_{\bar{\xi}}  \ . \label{eq:E.2}  
\end{equation}
Now to get the vector fields for Poincaré patch of AdS$_3$ we have to set $S=\bar{S}=0$ accordingly and it becomes:
\begin{equation}
 L_q= \frac{1}{2}(q+1)z\xi^q\partial_z+ \xi^{q+1}\partial_\xi -\frac{1}{2}q(q+1)z^2\xi^{q-1}\partial_{\bar{\xi}} \ . \label{eq:E.3}
\end{equation}
In the next part we write them in the $z,\tau,x$ and use the definitions of $\xi, \bar{\xi}$ to write $\partial_\xi=\frac{1}{2}(\partial_\tau-i\partial_x)$ and $\partial_{\bar{\xi}}=\frac{1}{2}(\partial_\tau+i\partial_x)$.

\section{Bulk Vector Fields \& Integral Curves}\label{appenF}

Let us now extend the boundary CFT Hamiltonian \eqref{eq:D.3} into the bulk by replacing the global Virasoro generators by the corresponding bulk AdS3 vector fields. We will work in the Euclidean patch.
\begin{align}
     &\hspace*{25mm} L_0 =-\frac{1}{2} i (\tau +i x) \partial_x +\frac{1}{2} (\tau +i x) \partial_{\tau } +\frac{1}{2} z \partial_z  \ , \notag
    \\
    &\hspace*{25mm} \bar{L}_0 =\frac{1}{2} i (\tau -i x) \partial_x +\frac{1}{2} (\tau -i x) \partial_{\tau } +\frac{1}{2} z \partial_z  \ , \notag
    \\ 
    L_q &=\frac{1}{2} (q+1) z (\tau +i x)^q \partial_z -\frac{1}{2} i \left((\tau +i x)^{q+1}+\frac{1}{2} (q+1) q z^2 (\tau +i x)^{q-1}\right)  \partial_x \notag
    \\
    &\hspace*{45mm} +\frac{1}{2}  \left((\tau +i x)^{q+1}-\frac{1}{2} q (q+1) z^2 (\tau +i x)^{q-1}\right) \partial_{\tau }  \ ,\notag
    \\
    \bar{L}_q &=\frac{1}{2} (q+1) z (\tau -i x)^q \partial_z +\frac{1}{2} i \left((\tau -i x)^{q+1}+\frac{1}{2} (q+1) q z^2 (\tau -i x)^{q-1}\right) \partial_x \notag
   &\hspace*{45mm} +\frac{1}{2} \left((\tau -i x)^{q+1}-\frac{1}{2} q (q+1) z^2 (\tau - i x)^{q-1}\right) \partial_{\tau } \ , \notag
    \\
   L_{-q} &=\frac{1}{2} (1-q) z (\tau +i x)^{-q} \partial_z -\frac{1}{2} i \left((\tau +i x)^{1-q}-\frac{1}{2} (1-q) q z^2 (\tau +i x)^{-q-1}\right)  \partial_x \notag
    \\
    &\hspace*{45mm} +\frac{1}{2}  \left((\tau +i x)^{1-q}+\frac{1}{2} q (1-q) z^2 (\tau +i x)^{-q-1}\right) \partial_{\tau }  \ , \notag
   \\
    \bar{L}_{-q} &=\frac{1}{2} (1-q) z (\tau -i x)^{-q} \partial_z +\frac{1}{2} i \left((\tau -i x)^{1-q}-\frac{1}{2} (1-q) q z^2 (\tau -i x)^{-q-1}\right) \partial_x \notag
    \\
   &\hspace*{45mm} +\frac{1}{2} \left((\tau -i x)^{1-q}+\frac{1}{2} q (1-q) z^2 (\tau - x)^{-q-1}\right) \partial_{\tau } \ . \label{eq:F.1}
\end{align}
Let us set $\alpha$ to $0$ and equate $\beta$ and $\gamma$ in the Floquet Hamiltonian \eqref{eq:D.3}. This greatly simplifies the computations while maintaining all of the dynamical richness.

Substituting \eqref{eq:F.1} into the Hamiltonian we can write down the explicit form of the integral curves using an intrinsic coordinate, denoted by $s \in \mathbb{R}$. The set of points that do not flow is represented by the fixed points in the bulk. The tangent equations of the curve produced by the bulk Hamiltonian give us:
\begin{align}
    \frac{dz(s)}{ds} &=-\frac{1}{2} \beta  z (q-1) (\tau -i x)^{-q}+(q+1) (\tau -i x)^q-(q-1) (\tau +i x)^{-q} \notag
    \\
    &\hspace*{85mm} +(q+1) (\tau +i x)^q \ , 
    \label{eq:F.2}
    \\
 \frac{d\tau(s)}{ds} &=-(\tau -i x)^{q-1} \left(q (q+1) z^2+2 (x+i \tau )^2\right)-(\tau +i x)^{q-1} \left(q (q+1) z^2-2 (\tau +i x)^2\right)
 \notag
 \\
& -(\tau +i x)^{-q-1} \left((q-1) q z^2-2 (\tau +i x)^2\right)
 -\frac{1}{4} \beta  (\tau -i x)^{-q-1} \left((q-1) q z^2-2 (x+i \tau )^2\right) \ ,
\label{eq:F.3}
\\
\frac{dx(s)}{ds} &=\frac{1}{4} i \beta  (\tau -i x)^{-q-1} \left((q-1) q z^2-2 (x+i \tau )^2\right)+(\tau -i x)^{q-1} \left(q (q+1) z^2-2 (x+i \tau )^2\right)
\notag
\\
& -(\tau +i x)^{-q-1} \left((q-1) q z^2+2 (\tau +i x)^2\right)-(\tau +i x)^{q-1} \left(q (q+1) z^2-2 (x-i \tau )^2\right) \ . \label{eq:F.4}
\end{align}
The fixed points are described by
\begin{equation}
    \, \frac{dz(s)}{ds}=0\ , \quad \frac{d\tau(s)}{ds}=0\ , \quad \frac{dz(x)}{ds}=0 \ .
    \label{eq:F.5}
    \end{equation}
    %
\section{Additional Fixed Points}\label{appenG}
In the main  section we have only focused on $\tau=0$ slice and we have also set $\alpha=0$. The integral curve equations, no general, yield more fixed points. Here, we present some representative cases to illustrate this point.

\begin{table}[h!]
\centering
\renewcommand{\arraystretch}{1.8} 
\setlength{\tabcolsep}{10pt} 
\begin{tabular}{|c|c|c|c|}
\hline
\multicolumn{4}{|c|}{\textbf{Solutions}} \\ \hline

$q$ & \textbf{Solution 1} & \textbf{Solution 2} & \textbf{Solution 3} \\ \hline
$q=2$ & 
\makecell{$z = \frac{\sqrt{x^2 - 4x^6}}{\sqrt{6x^4 - \frac{1}{2}}}$ \\ $\tau = x$} & 
\makecell{$z = \frac{\sqrt{x^2 - 4x^6}}{\sqrt{6x^4 - \frac{1}{2}}}$ \\ $\tau = -x$} & 
\makecell{$x = 0, \tau = -\frac{1}{3^{\frac{1}{4}}},$ , $z = \frac{\sqrt{2}}{3^{\frac{1}{4}}}$} \\ \hline

$q=3$ & 
\makecell{$z = \frac{2 \sqrt{x^2 - 64x^8}}{\sqrt{384x^6 - 3}}$ \\ $\tau = -\sqrt{3}x$} & 
\makecell{$x = 0, z = \frac{1}{2 \sqrt[6]{2}},$ \\ $\tau = \frac{1}{\sqrt[6]{2}}$} & 
\makecell{$x = \frac{\sqrt{3}}{2 \sqrt[6]{2}},$ , $z = \frac{1}{2 \sqrt[6]{2}},$ , $\tau = \frac{1}{2 \sqrt[6]{2}}$} \\ \hline

\end{tabular}
\caption{Fixed Points Solutions for $\tau \neq 0$ slices.}
\label{table:example}
\end{table}
Note that, the $x=0$ situations are interchangeable to the $\tau = 0$ cases, since in the Euclidean descriptions there is no distinction between them. 

\newpage
\bibliography{biblio}
\bibliographystyle{JHEP}

\end{document}